\documentclass{jfm}
\usepackage{graphicx}
\usepackage{epstopdf, epsfig}
\usepackage{amsmath}
\usepackage{mathtools}
\usepackage{graphicx}
\usepackage{siunitx}
\usepackage{caption}
\usepackage{subcaption}
\usepackage{natbib}
\usepackage{adjustbox}\usepackage{mathtools}
\usepackage{hhline}
\usepackage{array}\usepackage{makecell}
\usepackage{url} 
\usepackage{stackengine}\setstackEOL{\cr}
\setstackgap{S}{2ex}
\setcellgapes{3pt}
\usepackage[usenames]{color}

\newcommand{\beq}		{\begin{equation}}
\newcommand{\eeq}		{\end{equation}}

\newcommand{\ii}	    {\mathrm{i}}

\newcommand{\dd}	    {{\rm d}}

\DeclareMathAlphabet\mathbfcal{OMS}{cmsy}{b}{n}

\shorttitle{Wave packets and point vortices}
\shortauthor{N. Pizzo and R. Salmon}

\title{Particle description of the interaction between wave packets and point vortices}

\author{Nick Pizzo\aff{1}
  \corresp{\email{npizzo@ucsd.edu}} and 
  Rick Salmon\aff{1}}

\affiliation{\aff{1}Scripps Institution of Oceanography, University of California San Diego,
La Jolla, CA 92037, USA}

\begin{document}

\maketitle

\section*{Abstract}

This paper explores an idealized model of the ocean surface in which widely separated surface-wave packets and point vortices interact in two \textcolor{black}{horizontal} dimensions. We start with a Lagrangian which, in its general form,  depends on the \emph{fields} of wave action, wave phase, stream function, and two additional fields that label and track the \textcolor{black}{vertical component of} vorticity. By assuming that the wave action and vorticity are confined to infinitesimally small, widely separated regions of the flow, we obtain model equations that are analogous to, but significantly more general than, the familiar system consisting solely of point vortices. We analyze stable and unstable harmonic solutions, solutions in which wave packets eventually coincide with point vortices (violating our assumptions), and solutions in which the wave vector eventually blows up. \textcolor{black}{Additionally, we show that a wave packet induces a net drift on a passive vortex in the direction of wave propagation which is equivalent to Darwin drift. Generalizing our analysis to many wave packets and vortices,} we examine the influence of wave packets on an otherwise unstable vortex street and \textcolor{black}{show analytically, according to linear stability analysis, that the wave packet induced drift can stabilize the vortex street. The system is then numerically integrated for long times and an example is shown in which the configuration remains stable, which may be particularly relevant for the upper ocean.} 

\section{Introduction}

This paper explores an idealized model of the ocean surface in which widely separated surface-wave packets and point vortices interact in two \textcolor{black}{horizontal} dimensions. Each wave packet $p$ is defined by its location $\bold{x}_p(t)$, its wave action $\mathcal{A}_p$, and its wave vector $\bold{k}_p(t)$.  Each point vortex $i$ is defined by its location $\bold{x}_i(t)$ and its strength $\Gamma_i$. In reality, wave-breaking converts wave action into vorticity, and vorticity is destroyed by viscosity. However, in this initial study we consider only the ideal case, in which $\mathcal{A}_p$ and $\Gamma_i$ are conserved.  

The velocity field attached to the wave packets is dipolar; it is sometimes called `Bretherton flow'.  Wave packets advect the point vortices by their Bretherton flow.  Point vortices advect wave packets and other point vortices, and change the wave vector of the wave packets by refraction. For simplicity, we omit the interactions between wave packets, which are expected to be weak.

In \S 2 we derive the equations governing $\bold{x}_p(t)$, $\bold{k}_p(t)$, and $\bold{x}_i(t)$ from a Lagrangian which, in its general form,  depends on the \emph{fields} of wave action, wave phase, stream function, and two additional fields that label and track the \textcolor{black}{vertical component of} vorticity. In our application, the Lagrangian couples Whitham's Lagrangian for surface waves to the Langrangian for two-dimensional, incompressible flow. Coupling is achieved by replacing the `mean velocity' in the Doppler term of the dispersion relationship with the velocity field corresponding to the stream function of the vortical flow. We obtain our final equations by assuming that the wave action and vorticity are confined to infinitesimally small, widely separated regions of the flow. \textcolor{black}{To leading order, each wave packet induces a dipolar horizontal flow, and each vortex patch induces a monopolar flow.} In its general formulation \citep{Salmon2020}, the method applies to any type of wave and any type of mean flow in two or three dimensions.  It seems easier to apply than other, apparently equivalent methods that do not employ a Lagrangian.

In \S 3 we consider the system consisting of a single wave packet and a single \textcolor{black}{point} vortex. We analyze harmonic solutions in which the two particles move in circular orbits. \textcolor{black}{For these configurations,} we show that solutions in which the vortex orbit lies outside the orbit of the wave packet are stable, whereas solutions in which the vortex orbit lies inside that of the wave packet are unstable.  We also investigate solutions in which the vortex and wave packet eventually coincide, violating the assumption of our model, and solutions in which the wave vector grows without bound.

In \S 4 we consider the case of a wave packet encountering a pair of counter-rotating point vortices.  The highly symmetrical arrangement permits thorough analysis, which is confirmed by numerical solutions.  This solution is very similar to the one discussed by \citet{Buhler2005} and invites a comparison with their method of analysis. \textcolor{black}{We also show that, in the limit that the circulation of the vortices is much weaker than the wave action, the equations are equivalent to those diagnosing the motion of a particle in the presence of a uniformly translating cylinder. Following classical analysis \citep{Maxwell1870, Darwin1953} it is shown that the wave packet induces a net displacement on the vortices. }

In \S5 we study a solution in which $N>1$ wave packets are equidistant from, and symmetrically arranged about, a single vortex.  The wave packets circle the vortex at a uniform angular velocity, while the vortex remains stationary at the center of the pattern.

In \S 6 we generalize our system to be periodic in one dimension, and investigate the motion of a \textcolor{black}{periodic array of} weak point vortices in the presence of a \textcolor{black}{periodic array of} wave packets. We find asymptotic solutions \textcolor{black}{in which the wave packets induce a net drift on the vortices.} 

In \S 7, we use the analysis in \S 6 to investigate the linear stability of a vortex street in the presence of a wave packet. We find that the wave packet can change the stability of the vortex street. Numerical analysis demonstrates that vortex streets can be stable for long times in the presence of a wave packet. 

\S 8 concludes with an assessment of our results and their oceanographic implications.

\section{The equations of motion}
In this section we derive the equations governing a mixture of widely separated vortex patches and surface-wave packets.  In the wide-separation limit, the vortex patches correspond to point vortices and the wave packets correspond to `point dipoles.' We obtain our equations by coupling the Lagrangian for the wave field in the form proposed by \citet{whitham1965,Whitham1974} to the Lagrangian for two-dimensional incompressible flow representing the surface current.  We use the  Doppler term in the dispersion relation to couple the two Lagrangians together.  This appears to be a simple and powerful method for deriving equations governing the interactions between waves and mean flows. \textcolor{black}{Further details of the method are given by Salmon (2020).}

For the waves by themselves the Lagrangian proposed by Whitham is
\beq
L_w[\theta, \mathcal{A}]=\iiint dtd\mathbf{x}\; \left( \omega-\omega_r (\mathbf{k}) -\mathbf{U}\cdot \mathbf{k} \right)  \mathcal{A},
\label{01}
\eeq
where the integral is over time and the ocean surface; the frequency $\omega = -\theta_t$ and wave vector $\mathbf{k}=\nabla \theta$ are abbreviations for the derivatives of the wave phase $\theta(\mathbf{x},t)$;
\beq
\mathcal{A}=\frac{E}{\omega_r}
\label{02}
\eeq
is the wave action; $E$ is the wave energy per unit area; and $\omega_r (\mathbf{k})$ is the
prescribed relative frequency of the waves---the frequency measured in a reference frame moving at the mean flow velocity $\mathbf{U}(\mathbf{x},t)$.  
Our notation is $\mathbf{k}=(k,l)$,  $\mathbf{x}=(x,y)$, and $\nabla=(\partial_x,\partial_y)$. For surface waves, 
\beq
\textcolor{black}{\omega_r (\mathbf{k})= \sqrt{g\vert\mathbf{k}\vert}.}
\label{2a}
\eeq
\textcolor{black}{At this stage, we consider the mean flow to be prescribed.  For sake of completeness, Appendix A  provides a systematic derivation of \eqref{01} following Whitham's averaged-Lagrangian method.  Variations of $\mathcal{A}$ yield the dispersion relation,}
\beq
\textcolor{black}{\omega = \sqrt{g\vert\mathbf{k}\vert}
+\mathbf{U}\cdot \mathbf{k}.}
\label{03}
\eeq
From variations of $\theta$ we obtain
\begin{align}
\delta L_w[\theta, \mathcal{A}]&=\iiint dtd\mathbf{x}\; \left( 
-(\delta \theta)_t -\frac{\partial \omega_r}{\partial\mathbf{k}}\cdot \nabla (\delta \theta)
 -\mathbf{U}\cdot \nabla (\delta \theta) \right)  \mathcal{A}
\notag \\
 &=\iiint dtd\mathbf{x}\; \left( 
\mathcal{A}_t +\nabla \cdot [(\mathbf{c}_g+\mathbf{U}) \; \mathcal{A}]
\right)   \delta \theta,
\label{04}
\end{align}
where $\mathbf{c}_g(\mathbf{k})=\partial \omega_r / \partial \mathbf{k}$ is the relative group velocity.  Thus we obtain the action conservation
equation,
\beq
\mathcal{A}_t +\nabla \cdot [(\mathbf{c}_g+\mathbf{U}) \; \mathcal{A}] = 0.
\label{05}
\eeq

The Lagrangian for two-dimensional incompressible flow is
\beq
L_{m}[\alpha,\beta,\psi]=\iiint dt d\mathbf{x} \; H_0
\left( -\alpha \beta_t +\psi \frac{\partial(\alpha,\beta)}{\partial(x,y)} +\frac{1}{2} \nabla \psi \cdot \nabla \psi
 \right),
 \label{06}
\eeq
where the subscript $m$ stands for `mean flow', \textcolor{black}{and
\beq
 \frac{\partial(A,B)}{\partial(x,y)} \equiv [A,B] = A_x B_y - B_x A_y
\label{010}
\eeq
is the Jacobian, defined for any two functions $A(x,y)$ and $B(x,y)$}. The variables $\alpha(\mathbf{x},t)$, $\beta(\mathbf{x},t)$ and $\psi(\mathbf{x},t)$ represent averages over the constant depth $H_0$ to which wave-mean interactions occur.  We identify $H_0$ with the decay depth of the surface waves. We assume that the mean flow is depth-independent in this range.  Stationarity of $L_{m}$ implies
\begin{align}
&\delta \alpha: \;\;\;\;  \beta_t + [\psi,\beta]=0,
\label{07} \\
&\delta \beta: \;\;\;\; \alpha_t + [\psi,\alpha]=0,
\label{08} \\
&\delta \psi: \;\;\;\;   [\alpha,\beta]= \nabla^2 \psi, 
\label{09}
\end{align}
We see that $\alpha$ and $\beta$ are vorticity labels \textcolor{black}{in the following sense: First, by (\ref{07}) and (\ref{08}), they are conserved following the fluid motion, hence each fluid particle is identified by its two labels $(\alpha,\beta)$.  Second, by (\ref{09}), the vorticity in an arbitrary area of the flow is given by
\beq
\iint dxdy \nabla^2 \psi
= \iint dxdy \frac{\partial(\alpha,\beta)}{\partial(x,y)}
=\iint d\alpha d\beta,
\label{10a}
\eeq
where the integration is over the arbitrary area in physical space and the corresponding area in label space.}
Taking the time-derivative of \eqref{09} and using the Jacobi identity,
\beq
[A,[B,C]]+[B,[C,A]]+[C,[A,B]]=0,
\label{011}
\eeq
we obtain the vorticity equation,
\beq
\nabla^2 \psi _t+[\psi,\nabla^2 \psi]=0,
\label{012}
\eeq
for the mean flow by itself.

We couple $L_w$ to $L_{m}$ by replacing the mean velocity $\mathbf{U}$ in \eqref{01} with $\mathbf{u}_\psi\equiv(-\psi_y,\psi_x)$, and by assuming that the Lagrangian for the entire system is the sum,
\beq
L[\theta,\mathcal{A},\alpha,\beta,\psi]=L_w+L_{m}=\iiint dtd\mathbf{x}\; \left( 
-\theta_t\mathcal{A} -H_0 \alpha \beta_t   \right)     -\int dt \; H ,
\label{013}
\eeq
of \eqref{01} and \eqref{06}, where
\beq
H[\theta,\mathcal{A},\alpha,\beta,\psi]
=\iint d\mathbf{x} \; \left(
\omega_r  \mathcal{A} 
-H_0 \; \psi \frac{\partial(\alpha,\beta)}{\partial(x,y)} 
-\frac{H_0}{2} \nabla \psi \cdot \nabla \psi 
+\nabla \psi \times \mathbf{k} \mathcal{A} \right)
\label{014}
\eeq
is the Hamiltonian, and $(\theta,\mathcal{A})$ and $(\alpha,\beta)$ are canonical pairs.  Using \eqref{02}, \eqref{09} and integrations by parts to evaluate \eqref{014} we find that
\beq
H=\iint d\mathbf{x} \; \left( E
+\frac{H_0}{2} \nabla \psi \cdot \nabla \psi \right).
\label{015}
\eeq
Thus, \textcolor{black}{as expected}, our dynamics conserves the sum of the wave energy and the kinetic energy of the mean flow.  The equations corresponding to $\delta L=0$ are
\begin{align}
&\delta \mathcal{A}: \;\;\;\;  \omega \equiv - \theta_t =\omega_r(k,l,m)+\mathbf{u}_\psi\cdot \mathbf{k},
\label{016} \\
&\delta \theta: \;\;\;\; \mathcal{A}_t +\nabla \cdot [(\mathbf{c}_g+\mathbf{u}_\psi) \; \mathcal{A}] = 0,
\label{017} \\
&\delta \psi: \;\;\;\;  H_0[\alpha,\beta]=H_0\nabla^2 \psi- \nabla \times (\mathbf{k} \mathcal{A}) ,
\label{018} \\
&\delta \alpha: \;\;\;\;  \beta_t + [\psi,\beta]=0,
\label{019} \\
&\delta \beta: \;\;\;\; \alpha_t + [\psi,\alpha]=0,
\label{020}
\end{align}
where $\nabla \times (A,B) \equiv B_x-A_y$ \textcolor{black}{will be our notation for the vertical component of the curl of a horizontal vector.}
By the Jacobian identity, \eqref{018}-\eqref{020} imply
\beq
q_t + [\psi,q]=0,
\label{021}
\eeq
where
\beq
q = H_0\nabla^2 \psi-\nabla \times \mathbf{p},
\label{022}
\eeq
and 
\beq
\mathbf{p} = \mathbf{k} \mathcal{A}
\label{023}
\eeq
is the pseudomomentum.   The wave action equation \eqref{04} is unchanged, but now
$\mathbf{U}=\mathbf{u}_\psi$. \textcolor{black}{That is, the previously arbitrary mean flow is now specifically identified with the velocity field $(-\psi_y,\psi_x)$ induced by the point vortices and wave packets.}

The most interesting effect of the coupling and summation of Lagrangians is the generalization of \eqref{012} to \eqref{021}-\eqref{022}.  By these equations, the quantity $H_0\nabla^2 \psi-\nabla \times \mathbf{p}$ is conserved following the mean motion of fluid particles.  Consider waves propagating into a region of fluid that is initially at rest.  Before the arrival of the waves, $\nabla^2 \psi=\mathbf{p}=0$, and hence
\beq
H_0\nabla^2 \psi = \nabla \times \mathbf{p}.
\label{024}
\eeq
By \eqref{021}, \eqref{024} applies at \emph{all} times, including when waves are present.  Equation \eqref{024} is a concise definition of Bretherton flow, the flow generated by a wave packet in a formerly quiescent fluid.  If wave breaking destroys the pseudomomentum $\mathbf{p}$ before the broad mean flow represented by $\psi$ has time to react, then real, actual, vorticity is created and remains behind after the remaining wave energy propagates away.

Taking the gradient of the dispersion relation \textcolor{black}{ (\ref{03}) and using 
$\nabla \omega= - \nabla \theta_t= -\mathbf{k}_t$,}
we obtain the refraction equation
\beq
\frac{\partial \mathbf{k}}{\partial t}
+\left( (\mathbf{c}_g+\mathbf{U})\cdot \nabla \right)\mathbf{k}
=-k \nabla U-l\nabla V.
\label{025}
\eeq
The refractive change in $\mathbf{k}$ predicted by \eqref{025} causes a change in $\omega_r(\mathbf{k})$ that can be determined from \eqref{2a}. 
If the waves do not break, then the action $\mathcal{A}=E/\omega_r(\mathbf{k})$ is conserved.   If $\omega_r(\mathbf{k})$ increases, then the wave energy $E$ must also increase to keep their ratio constant.  We anticipate that wave-vector stretching, which increases $\vert \mathbf{k} \vert$ and hence $\omega_r$, is \emph{typical} for the same reason that fluid particles typically move apart, and hence wave-mean interactions typically transfer energy from surface currents to waves.
On the other hand, wave breaking always transfers energy from waves to currents.

Now we specialize the dynamics \eqref{013}-\eqref{014} to the case in which the vorticity and wave action are concentrated at widely separated points.  This specialization is motivated by a desire to produce equations amenable to analytical and numerical solution.
We assume that the mean flow vorticity consists solely of point vortices.  Then
\beq
[\alpha,\beta]=\sum_i \Gamma_i \delta\left(\mathbf{x}-\mathbf{x}_i(t)\right),
\label{026}
\eeq
where $\mathbf{x}_i(t)$ is the location at time $t$ of a point vortex with strength $\Gamma_i$.
The subscripts $i$ replace the continuous vorticity labels $\alpha$ and $\beta$.
The Hamiltonian \eqref{014} becomes
\beq
H[\theta,A,\mathbf{x}_i,\psi]
=-\sum_i H_0\Gamma_i \psi(\mathbf{x}_i(t))
+\iint d\mathbf{x} \; \left(
\omega_r(\theta_x,\theta_y)  \mathcal{A}
 -\frac{H_0}{2} \nabla \psi \cdot \nabla \psi  +[\psi,\theta] \mathcal{A} \right).
 \label{027}
\eeq
To fully convert from $(\alpha,\beta)$ to $\mathbf{x}_i$ we must transform the term
\beq
\iiint d\mathbf{x} dt  \; \alpha \beta_t 
\label{028}
\eeq
in \eqref{013}.  It becomes
\begin{align}
&\iiint dx dy dt \; \alpha \frac{\partial (x,y,\beta)}{\partial(x,y,t)}
=\iiint d\alpha d\beta d\tau \; \alpha \frac{\partial (x,y,\beta)}{\partial(\alpha,\beta,\tau)}
\notag \\
&=\iiint d\alpha d\beta d\tau \; -x \frac{\partial (\alpha,y,\beta)}{\partial(\alpha,\beta,\tau)}
=\iiint d\alpha d\beta d\tau \; x\frac{\partial y}{\partial \tau}
=\int dt \sum_i \Gamma_i x_i \frac{dy_i}{dt}.
\label{029}
\end{align}
Thus, when point vortices replace continuous vorticity, the Lagrangian \eqref{013} becomes
\begin{align}
L[\theta,A,\mathbf{x}_i,\psi]=
\iiint d\mathbf{x} dt & \; \left(
-\theta_t\mathcal{A} -\omega_r(\theta_x,\theta_y)  \mathcal{A}
 +\frac{H_0}{2} \nabla \psi \cdot \nabla \psi  -[\psi,\theta] \mathcal{A} \right)
\notag \\
&+\int dt \left( -\sum_i H_0\Gamma_i x_i \frac{dy_i}{dt}
+ \sum_i H_0\Gamma_i \psi(\mathbf{x}_i(t))  \right).
\label{030}
\end{align}
Instead of \eqref{018} we now have
\beq
\delta \psi: \;\;\;\; H_0\nabla^2 \psi = 
\sum_i H_0\Gamma_i  \delta\left(\mathbf{x}-\mathbf{x}_i(t)\right)+[\mathcal{A},\theta],
\label{031}
\eeq
with solution
\beq
\psi(\mathbf{x},t)=\frac{1}{2\pi}\sum_i \Gamma_i  \ln \vert\mathbf{x}-\mathbf{x}_i(t)\vert
+\psi_w(\mathbf{x},t),
\label{032}
\eeq
where (suppressing the time-dependence)
\beq
\psi_w(\mathbf{x})=\frac{1}{H_0}\iint d\mathbf{x}' \; \rho(\mathbf{x}') \frac{1}{2\pi} \ln \vert \mathbf{x} - \mathbf{x}' \vert
\label{033}
\eeq
and
\beq
\rho =[\mathcal{A},\theta] = \nabla \mathcal{A} \times \mathbf{k}.
\label{034}
\eeq

We now assume that the wave field consists solely of isolated wave packets.
The stream function field generated by a single wave packet at $\mathbf{x}_p$ is given by \eqref{033} and \eqref{034} with $\mathbf{k}=\mathbf{k}_p$, where $\mathbf{k}_p(t)$ is the wave vector associated with the wave packet.  We assume that $\mathbf{k}_p$ depends only on time;  its variation within the wave packet is assumed negligible.  
The integration in \eqref{033} is over the area of the wave packet, the region of the flow in which $\mathcal{A}\neq 0$.  In Appendix B we show that, far from $\mathbf{x}_p$,  the streamfunction generated by a wavepacket at $\mathbf{x}_p$ takes the form of a dipole,
\beq
\psi(\mathbf{x})= 
 \frac{1}{2 \pi H_0} \frac {(\mathbf{x}-\mathbf{x}_p) \times \mathbf{k}_p}{\vert \mathbf{x}-\mathbf{x}_p \vert^2} 
\mathcal{A}_p,
\label{035}
\eeq
where $\mathcal{A}_p = \iint d\mathbf{x} \mathcal{A}$ is the total action of the wave packet.  

The streamfunction response to many point vortices and many wave packets is clearly
\beq
\psi(\mathbf{x},t)=\sum_i \Gamma_i  \psi_m(\mathbf{x},\mathbf{x}_i)
+ \sum_p \mathcal{A}_p \psi_d(\mathbf{x},\mathbf{x}_p,\mathbf{k}_p),
\label{036}
\eeq
where
\beq
\psi_m(\mathbf{x},\mathbf{x}_i)\equiv
\frac{1}{2\pi}\ln \vert \mathbf{x}-\mathbf{x}_i(t)\vert
\label{037}
\eeq
is the response to a monopole at $\mathbf{x}_i$, and
\beq
\psi_d(\mathbf{x},\mathbf{x}_p,\mathbf{k}_p)\equiv
\frac{1}{2\pi H_0}\frac {(\mathbf{x}-\mathbf{x}_p) \times \mathbf{k}_p}{\vert \mathbf{x}-\mathbf{x}_p \vert^2}
\label{038}
\eeq
is the response to a dipole with wavevector $\mathbf{k}_p$ at $\mathbf{x}_p$.  The constants $\Gamma_i$ and $\mathcal{A}_p$ measure the strength of the monopole and the dipole, respectively.  $\mathcal{A}_p$ is always positive but $\Gamma_i$ can have either sign.  Until dissipation occurs $\mathcal{A}_p$ and $\Gamma_i$ remain constant.  

Since our aim is to produce a Lagrangian that depends only on the point vortex locations $\mathbf{x}_i(t)$, the wave packet locations $\mathbf{x}_p(t)$, and their wave vectors $\mathbf{k}_p(t)$, we must transform all of the terms in \eqref{030}.   If we integrate the first term in \eqref{030} over the $p$-th wave packet, we obtain
\begin{align}
&-\iiint d\mathbf{x} dt \; \theta_t\mathcal{A}
=\iiint d\mathbf{x} dt \; \theta \mathcal{A}_t
=-\iiint d\mathbf{x} dt \; \theta \; \frac{d\mathbf{x}_p}{dt}\cdot \nabla \mathcal{A}
\notag\\
&=-\int dt \; \frac{d\mathbf{x}_p}{dt} \cdot \iint d\mathbf{x} \; \theta \nabla \mathcal{A}
=\int dt \; \frac{d\mathbf{x}_p}{dt} \cdot \iint d\mathbf{x} \;   \mathcal{A} \nabla \theta
= \int dt \; \frac{d\mathbf{x}_p}{dt} \cdot \mathbf{k}_p   \mathcal{A}_p,
\label{039}
\end{align}
where we have used integrations by parts and the relation
\beq
\left( \frac{\partial}{\partial t} + \frac{d\mathbf{x}_p}{dt} \cdot \nabla \right) \mathcal{A}(\mathbf{x},t)=0,
\label{040}
\eeq
which follows from the definition of $\mathbf{x}_p(t)$: $d\mathbf{x}_p(t)/dt$ is the velocity of the wave envelope.  
The second term in \eqref{030} becomes
\beq
-\int dt  \; \omega_r(\mathbf{k}_p)  \mathcal{A}_p.
\label{041}
\eeq
The three terms in \eqref{030} containing $\psi$ combine as
\begin{align}
& \iiint dtd\mathbf{x} \; \left(
\frac{H_0}{2} \nabla \psi \cdot \nabla\psi -[\psi,\theta] \mathcal{A} \right)
+  \int dt \sum_i H_0\Gamma_i   \psi(\mathbf{x}_i(t),t)
\notag \\
&= \iiint dtd\mathbf{x} \left(
-\frac{H_0}{2} \; \psi \nabla^2 \psi -[\psi,\theta] \mathcal{A} 
 + \; \sum_i H_0\Gamma_i \psi \delta(\mathbf{x}-\mathbf{x}_i) \right)
\notag \\
&=\iiint dtd\mathbf{x} \; \left(
-\frac{H_0}{2}  \psi \nabla^2 \psi +[\mathcal{A},\theta] \psi
+ \psi \left( H_0\nabla^2 \psi - [\mathcal{A},\theta] \right) \right)
\notag \\
&= \frac{H_0}{2} \iiint dtd\mathbf{x} \;  \psi \nabla^2 \psi,
\label{042}
\end{align}
where we have used \eqref{031}.  

Our final step is to substitute \eqref{036} back into the Lagrangian, removing its dependence on $\psi$.  
The last integral in \eqref{042} becomes
\begin{align}
\iint d\mathbf{x} \; \psi \nabla^2 \psi =
\iint d\mathbf{x} \; \bigg[
&\left( \sum_i \Gamma_i \psi_m(\mathbf{x},\mathbf{x}_i) +
\sum_p \mathcal{A}_p \psi_d(\mathbf{x},\mathbf{x}_p,\mathbf{k}_p) \right) \times
\notag \\
&\nabla^2
\left( \sum_j \Gamma_j \psi_m(\mathbf{x},\mathbf{x}_j) +
\sum_q \mathcal{A}_q \psi_d(\mathbf{x},\mathbf{x}_q,\mathbf{k}_q) \right)    \bigg].
\label{043}
\end{align}
We simplify \eqref{043} by neglecting the dipole-dipole interactions, which are expected to be weak: The velocity field associated with the monopoles falls off like $1/r$, whereas the velocity field associated with Bretherton dipoles falls off like $1/r^2$.
Dropping these terms from \eqref{043} gives us
\begin{align}
&\iint d\mathbf{x} \; \psi \nabla^2 \psi 
\approx   \iint d\mathbf{x}
 \sum_i \Gamma_i \nabla^2  \psi_m(\mathbf{x},\mathbf{x}_i)
\left( \sum_j \Gamma_j \psi_m(\mathbf{x},\mathbf{x}_j) +2 \sum_p \mathcal{A}_p \psi_d(\mathbf{x},\mathbf{x}_p,\mathbf{k}_p) \right)
\notag \\
&=\iint d\mathbf{x} \sum_i \Gamma_i \delta (\mathbf{x}-\mathbf{x}_i)
\left( \sum_j \Gamma_j \psi_m(\mathbf{x},\mathbf{x}_j) 
+2 \sum_p \mathcal{A}_p\psi_d(\mathbf{x},\mathbf{x}_p,\mathbf{k}_p) \right) 
\notag \\
&= \sum_i \Gamma_i
\left( \sum_j \Gamma_j \psi_m(\mathbf{x_i},\mathbf{x}_j) +2 \sum_p \mathcal{A}_p \psi_d(\mathbf{x_i},\mathbf{x}_p,\mathbf{k}_p) \right). 
\label{044}
\end{align}

Putting all this together, we obtain the Lagrangian 
\beq
L[ \mathbf{x}_i,\mathbf{x}_p,\mathbf{k}_p]=
\int dt \left(
\sum_p \mathcal{A}_p  \mathbf{k}_p \cdot \dot{\mathbf{x}}_p 
- \sum_i H_0\Gamma_i x_i \dot{y}_i 
-H[\mathbf{x}_i,\mathbf{x}_p,\mathbf{k}_p] \right),
\label{045}
\eeq
where
\begin{align}
H[\mathbf{x}_i,\mathbf{x}_p,\mathbf{k}_p]
&=
\sum_p \mathcal{A}_p \; \omega_r(\mathbf{k}_p)
-  \frac{H_0}{2\pi}
\sum_i \sum_{j>i} \Gamma_i \Gamma_j   \ln \vert \mathbf{x}_i -\mathbf{x}_j \vert
\notag \\
&-\frac{1}{2\pi} \sum_i \sum_p \Gamma_i \mathcal{A}_p 
\frac{(\mathbf{x}_i-\mathbf{x}_p) \times \mathbf{k}_p} {\vert \mathbf{x}_i-\mathbf{x}_p \vert ^2}
\label{046}
\end{align}
is the Hamiltonian.
For every wave packet there are two canonical pairs, $(x_p,k_p)$ and $(l_p,y_p)$, and for every point vortex there is one canonical pair, $(x_i,y_i)$.  Again, $\Gamma_i$ and $\mathcal{A}_p$ are constants.  The Hamiltonian \eqref{046} contains $\Gamma \Gamma$ terms and $\Gamma \mathcal{A}$ terms.  If we had not dropped the dipole/dipole interactions it would also contain $\mathcal{A} \mathcal{A}$ terms.

We remark that it is generally quite \emph{wrong} to substitute an equation resulting from the variational principle back into the Lagrangian.   If, for example, we substitute the dispersion relation back into \eqref{01}, the Lagrangian vanishes.  However, it \emph{is} legitimate to use the equation obtained by varying a particular field to eliminate that same field from the Lagrangian; \textcolor{black}{see Appendix C.}  Thus it is legal to use \eqref{031} to eliminate $\psi$ from \eqref{030}.

The equations corresponding to \eqref{045}-\eqref{046} are
\begin{align}
\delta \mathbf{k}_p: \;\;\;\;
&\dot{\mathbf{x}}_p=\frac{1}{\mathcal{A}_p} \frac{\partial H}{\partial \mathbf{k}_p}
=\mathbf{c}_g(\mathbf{k}_p)
+\mathbf{U}_m(\mathbf{x}_p),
\label{047} \\
\delta \mathbf{x}_p: \;\;\;\;
&\dot{\mathbf{k}}_p=-\frac{1}{\mathcal{A}_p} \frac{\partial H}{\partial \mathbf{x}_p}
=-k_p \nabla U_m(\mathbf{x}_p)-l_p\nabla V_m(\mathbf{x}_p),
\label{048} \\
\delta \mathbf{x}_i: \;\;\;\;
&\dot{\mathbf{x}}_i=\frac{1}{\Gamma_i} 
\left( \frac{\partial H}{\partial y_i}, -\frac{\partial H}{\partial x_i} \right)
=\mathbf{U}_m(\mathbf{x}_i)+\mathbf{U}_d(\mathbf{x}_i),
\label{049}
\end{align}
where
\begin{align}
\mathbf{U}_m(\mathbf{x})=(U_m(\mathbf{x}),V_m(\mathbf{x}))
&=\sum_i \Gamma_i     \left( 
-\frac{\partial \psi_m}{\partial y}(\mathbf{x},\mathbf{x}_i), \;
\frac{\partial \psi_m}{\partial x}(\mathbf{x},\mathbf{x}_i) \right)
\notag \\
&=\frac{1}{2\pi} \sum_i \Gamma_i \frac{(y_i-y,x-x_i)}{\vert \mathbf{x}_i-\mathbf{x} \vert ^2}
\label{050}
\end{align}
is the velocity field induced by the point vortices, and
\beq
\mathbf{U}_d(\mathbf{x})=\sum_p \mathcal{A}_p    \left( 
-\frac{\partial \psi_d}{\partial y}(\mathbf{x},\mathbf{x}_p,\mathbf{k}_p), \;
\frac{\partial \psi_d}{\partial x}(\mathbf{x},\mathbf{x}_p,\mathbf{k}_p) \right)
\label{051}
\eeq
is the velocity field induced by the wave packets.  The total velocity is 
$\mathbf{U}(\mathbf{x})=\mathbf{U}_m(\mathbf{x})+\mathbf{U}_d(\mathbf{x})$.
In our approximation, the wave packets talk to point vortices but not to one another, while the point vortices talk to both point vortices and wave packets.  We can add the missing physics if necessary; it would, for example, add the term $\mathbf{U}_d(\mathbf{x}_p)$ to \eqref{047}.

\textcolor{black}{
Equations (\ref{047}-\ref{049}) are the fundamental equations of our model.  If we were to regard $\mathbf{U}_m$ as a prescribed mean flow, then (\ref{047}) and (\ref{048}) would be the standard equations of ray theory (e.g. B\"{u}hler 2014).  Similarly, if we omit $\mathbf{U}_d$, then (\ref{049}) is the standard equation of point vortex dynamics \textcolor{black}{\citep{Kirchhoff1883}}.
The new feature of our derivation is that $\mathbf{U}_m$ is not prescribed, but rather is determined by the locations of the point vortices. Similarly, the dipolar velocity field of the wave packets is not dropped, but rather contributes to the advection of the point vortices.  Again, if the relatively weak interactions between the wave packets had not been dropped, then $\mathbf{U}_d$ would also appear in (\ref{047}) and (\ref{048}). \citet{tchieu2012} consider a system consisting solely of interacting point dipoles.  In our context, their system corresponds to adding dipole-dipole interactions but completely omitting the point vortices. }

\textcolor{black}{
The derivation of (\ref{047}-\ref{049}) from a Lagrangian guarantees that our dynamics maintains important conservation laws.  The conservation of energy (\ref{046}) corresponds to the time-translation symmetry of (\ref{045}-\ref{046}). The conservation of momentum,}
\beq
\mathbfcal{M}=\sum_p \mathcal{A}_p \mathbf{k}_p + H_0\sum_i\Gamma_i(y_i,-x_i).
\label{052}
\eeq
\textcolor{black}{corresponds to space-translation symmetry and} is proved by considering variations of the form
\beq
\delta \mathbf{x}_i =\delta \mathbf{x}_p= \boldsymbol{\epsilon}(t),
\label{053}
\eeq
where $\boldsymbol{\epsilon}(t)$ is an arbitrary infinitesimal vector.  If we think of the interactions between the dipoles and point vortices as the sum of pair interactions between each dipole/vortex pair, then pairwise conservation of \eqref{052} shows that the refraction of wave packet $p$ (i.e. the change in $\mathbf{k}_p$) caused by vortex $i$ is accompanied by a change in the position of vortex $i$.  \citet{buhler2003} refer to this as `remote recoil.'  
Conservation of \eqref{052} also governs wave breaking in the following sense.  If the $p$-th wave packet is completely destroyed by wave breaking, then $\mathcal{A}_p$ is suddenly replaced by two counter-rotating vortices with a dipole moment equal to $\Gamma D$ where $D$ is the separation between counter-rotating vortices of strength $\pm \Gamma$.  See also \citet{Buhler2005, Buhler2001}, and \citet{Buhler2014}. Our dynamics also conserves the angular momentum,
\beq
\mathcal{L}=
\sum_p \mathcal{A}_p  (\mathbf{k}_p \times \mathbf{x}_p) 
+ \frac{H_0}{2} \sum_i \Gamma_i  (x_i^2 + y_i^2),
\label{054}
\eeq
which can be proved by considering variations of the form
$(x_i+iy_i) \rightarrow (x_i+iy_i) e^{i\delta \theta(t)}$, 
$(x_p+iy_p) \rightarrow (x_p+iy_p) e^{i\delta \theta(t)}$,
and $(k_p+il_p) \rightarrow (k_p+il_p) e^{i\delta \theta(t)}$, where $\delta \theta(t)$ is an infinitesimal angle.

\textcolor{black}{
Our primary motivation in deriving (\ref{047}-\ref{049}) is to obtain relatively simple equations, amenable to analytic and numerical solution, governing the interactions between ocean surface waves and the vorticity created by breaking waves. One could avoid the assumption of widely spaced wave packets and vortex patches by solving the more general equations (\ref{016}-\ref{025}) as coupled equations for the \emph{fields} of $\mathcal{A}(\mathbf{x},t)$, $\mathbf{k}(\mathbf{x},t)$, and $\psi(\mathbf{x},t)$.  However, even these more general equations are idealized in the sense that they embody our treatment of the vertical dimension.  In reality, the vorticity associated with a surface wave packet resides in a horseshoe-shaped vortex tube whose surface manifestation is the vortex pair represented by our dipole.  The three-dimensional structure of this vortex tube is \emph{locally} important, but only the nearly vertical portions of the vortex tube induce a significant surface flow far from the tube.  The arbitrary constant depth $H_0$ to which the tube's contribution extends is an artificial component of our model and could easily be absorbed into other parameters.  We prefer to retain it as a constant reminder of the very idealized nature of our model. Models, to be useful, must be much simpler than reality.
}

\citet{Onsager1949} considered the equilibrium statistical mechanics of a system of point vortices.  Our system reduces to Onsager's system when no waves are present ($\mathcal{A}_p\equiv 0$).   Our phase space is larger than the one considered by Onsager because it contains dimensions corresponding to the wave packet locations $\mathbf{x}_p$ and their wave vectors $\mathbf{k}_p$.  However, the difference is not merely a matter of extra dimensions.  In Onsager's problem the volume of the phase space is \emph{finite}, because the point vortices are confined to a box.  In our problem the phase space has infinite volume because $-\infty<\mathbf{k}_p <\infty$.  We therefore expect an ultraviolet catastrophe in which energy spreads to ever larger $\vert \mathbf{k}_p \vert$ by the process of wave vector stretching.  If wave vector stretching increases the first term in \eqref{046}, as would be the case for surface waves, this increase must be compensated by a decrease in the other two terms.  

Our method is easily adapted to other types of waves and mean flows.  For example, to investigate internal waves interacting with a quasigeostrophic mean flow, we need only replace \eqref{03} with the dispersion relation for internal waves, and \eqref{06} with the Lagrangian for quasigeostrophic flow.  This approach offers advantages of simplicity and transparency over the more formal approaches followed by \citet{Buhler2005}, \citet{Wagner2015}, and \citet{Salmon2016}.  For many further details, see \citet{Salmon2020}. In the remainder of this paper we investigate the dynamics \eqref{047}-\eqref{049}.


\section{1 vortex, 1 wave packet}
We begin by considering the system consisting of a single vortex of strength $\Gamma$ located at $\mathbf{x}(t)$, and a single wavepacket with action $\mathcal{A}$ and wave vector $\mathbf{k}(t)$ located at $\mathbf{x}(t)+\boldsymbol{\xi}(t)$. This system exhibits a much more complicated range of behavior than the more familiar system consisting of two point vortices.
The Langrangian \eqref{045} takes the form
\beq
L[ \mathbf{x},\boldsymbol{\xi},\mathbf{k}]=
\int dt \left(
\mathcal{A} \; \mathbf{k} \cdot ( \dot{\boldsymbol{\xi}} +\dot{ \mathbf{x}})
- H_0 \Gamma x \dot{y} 
-\mathcal{H}(\boldsymbol{\xi},\mathbf{k}) \right),
\label{055}
\eeq
with Hamiltonian
\beq
\mathcal{H}(\boldsymbol{\xi},\mathbf{k})= \mathcal{A} \sqrt{g \vert \mathbf{k} \vert}-
\frac{1}{2\pi} \Gamma \mathcal{A} \frac{\mathbf{k} \times \boldsymbol{\xi}}
{\vert \boldsymbol{\xi} \vert^2}.
\label{056}
\eeq
The equations of motion become
\begin{align}
&\delta x : \;\;\;\; 
\frac{d}{dt}(\mathcal{A}k+H_0 \Gamma y)=0,
\label{057} \\
&\delta y : \;\;\;\;
\frac{d}{dt}(\mathcal{A}l-H_0 \Gamma x)=0,
\label{058} \\
&\delta \mathbf{k} : \;\;\;\;
\frac{d}{dt} (\boldsymbol{\xi}+\mathbf{x}) =\mathbf{c}_g(\mathbf{k})+
\frac{\Gamma}{2\pi \vert \boldsymbol{\xi} \vert^2}(-\eta,\xi),
\label{059} \\
&\delta \xi : \;\;\;\;
\frac{dk}{dt}=\frac{\Gamma}{2\pi \vert \boldsymbol{\xi} \vert^4}
\left( l(\xi^2-\eta^2)-2k \xi \eta \right),
\label{060} \\
&\delta \eta : \;\;\;\;
\frac{dl}{dt}=\frac{\Gamma}{2\pi \vert \boldsymbol{\xi} \vert^4}
\left( k(\xi^2-\eta^2)+2l \xi \eta \right),
\label{061}
\end{align}
where $\boldsymbol{\xi}=(\xi,\eta)$. \textcolor{black}{We simplify notation by taking $g=1$ and choosing a characteristic wavenumber $k_0=1$ so that $\omega_0^2=gk_0=1$. We also assume $H_0=k_0^{-1}=1$, while we take $\Gamma = 2\pi \omega_0 k_0^{-2}=2\pi$ and $\mathcal{A}=2\pi\omega_0k_0^{-5} = 2\pi$.} Then the Hamiltonian \eqref{056} becomes  
\beq
\mathcal{H} = 2\pi\left((k^2+l^2)^{1/4}- \frac{k\eta - l \xi}{\xi^2+\eta^2}\right).
\label{062}
\end{equation}

The system \eqref{057}-\eqref{061} conserves the energy \eqref{062}, the
angular momentum 
\beq
\mathcal{L} = 2\pi (\bold{k}\times (\bold{x}+\boldsymbol{\xi})) + \frac{2\pi}{2}(x^2+y^2),
\label{063}
\eeq
(cf. \ref{054}), and the momentum  $\mathbfcal{M}\equiv (\mathcal{M}_x,\mathcal{M}_y)$ (cf. \ref{052}), where 
\beq
\mathcal{M}_x = 2\pi (k+  y); \quad \mathcal{M}_y=2\pi(l-  x).
\label{064}
\eeq
We use the conserved momenta \eqref{064} to eliminate the variables $(x,y)$ in favor of $(k,l,\xi,\eta)$.  The resulting system conserves the energy \eqref{062} and the quantity
\beq
\mathcal{R}_0 \equiv
\frac{1}{2}\left(\left(\frac{\mathcal{M}_x}{2\pi}\right)^2+\left(\frac{\mathcal{M}_y}{2\pi}\right)^2\right)-\frac{  \mathcal{L}}{2\pi} = \frac{1}{2}(k^2+l^2)-  (k\eta - l\xi)
\label{065}
\eeq
obtained by eliminating $(x,y)$ between \eqref{063} and \eqref{064}.
We also define 
\beq
\mathcal{H}_0 = \frac{\mathcal{H}}{2\pi}.
\label{066}
\eeq
The reduced dynamics takes the form of four coupled ordinary differential equations,
\beq
\dot{\xi}= -\frac{ k(\xi^2-\eta^2)+2l\xi \eta}{(\xi^2+\eta^2)^2}+\frac{k}{2(k^2+l^2)^{3/4}}- \frac{  \eta }{\xi^2 + \eta^2},
\label{067}
\eeq
\beq
\dot{\eta} = \frac{ l(\xi^2-\eta^2)-2k\xi \eta}{(\xi^2+\eta^2)^2}+\frac{l}{2(k^2+l^2)^{3/4}}+\frac{  \xi }{\xi^2 + \eta^2},
\label{068}
\eeq
\beq
\dot{k} =  \frac{ l(\xi^2-\eta^2)-2k\xi \eta}{(\xi^2+\eta^2)^2},
\label{069}
\eeq
\beq
 \dot{l} = \frac{ k(\xi^2-\eta^2)+2l\xi \eta}{(\xi^2+\eta^2)^2}.
\label{070}
\eeq
with the  two conserved quantities, \eqref{065} and \eqref{066}.

Define 
\beq
k+\ii l = \kappa e^{\ii\phi}, \quad \xi+ \ii \eta = a e^{\ii\theta}.
\label{072}
\eeq
We shall obtain a single, closed equation for the wavenumber magnitude $\kappa(t)$.
First, using \eqref{062} and \eqref{066}, we obtain an expression for $a^2$ in terms of $\kappa$,
\beq
a^2= \frac{\mathcal{R}_0-\frac{1}{2}\kappa ^2}{\mathcal{H}_0-\sqrt{\kappa}}.
\label{073}
\eeq
Then using (\ref{062}) we obtain the constraint
\beq
\sin (\phi-\theta) = \frac{a}{  \kappa} \left(\mathcal{H}_0-\sqrt{\kappa}\right)
\label{074}
\eeq
on the phases.
From (\ref{069}) and (\ref{070}), we find an evolution equation for $\phi$, namely
\beq
\dot{\phi}=\frac{-l\dot{k}+k\dot{l}}{k^2+l^2}=\frac{1}{a^2} \cos 2(\phi-\theta).
\label{075}
\eeq
Equations (\ref{069}) and (\ref{070}) also imply
\beq
\dot{\kappa} = \frac{\dot{k}k+\dot{l}l}{\sqrt{k^2+l^2}} = \frac{\kappa}{a^2} \sin 2(\phi-\theta). 
\label{076}
\eeq
Combining (\ref{075}) and (\ref{076}), we obtain
\beq
\dot{\kappa}^2+\kappa^2 \dot{\phi}^2 = \frac{\kappa^2}{a^4} = \kappa^2 \left(\frac{\mathcal{H}_0-\sqrt{\kappa}}{\mathcal{R}_0-\frac{1}{2}\kappa ^2}\right)^2.
\label{077}
\eeq
Our final step is to eliminate $\dot{\phi}$ to arrive at equation involving only $\dot{\kappa}$ and $\kappa$. We use the identity
\beq
\cos 2(\phi - \theta) =1- 2\sin^2 (\phi-\theta) = 1-\frac{2 a^2}{\kappa^2} (\mathcal{H}_0-\sqrt{\kappa})^2,
\label{078}
\eeq
where the last substitution is via (\ref{073}) and (\ref{074}).
Then by \eqref{073} and \eqref{078} we have
\beq
\dot{\phi} = \frac{\mathcal{H}_0-\sqrt{\kappa}}{\mathcal{R}_0-\frac{1}{2}\kappa ^2}-\frac{2}{\kappa^2} (\mathcal{H}_0-\sqrt{\kappa})^2.
\label{079}
\eeq
Substituting \eqref{079} back into \eqref{077} we obtain the closed evolution equation
\beq
\frac{1}{2}\dot{\kappa}^2+\Pi(\kappa)=0
\label{080}
\eeq
for $\kappa(t)$, where 
\beq
\Pi(\kappa)=\frac{1}{2}\left(\frac{\kappa (\mathcal{H}_0-\sqrt{\kappa})}{(\mathcal{R}_0-\frac{1}{2}\kappa^2)}\right)^2\left(-1+ \left(1-\frac{2}{ \kappa^2}(\mathcal{R}_0-\frac{1}{2}\kappa^2)(\mathcal{H}_0-\sqrt{\kappa})\right)^2 \right).
\label{081}
\eeq
Equation (\ref{080}) takes the form of a particle moving in a potential $\Pi(\kappa)$. This permits a qualitative analysis of system behavior based on the form of \eqref{081}. Solutions may be written out in implicit form, as  in \citet{Tur2017}, but a qualitative analysis offers better physical insight.

\subsection{Circular motion}
We begin by seeking solutions that exhibit simple harmonic motion. Thus we take $\dot{\kappa}=0$ and look for the  $\kappa_i$ that satisfy
\beq
\Pi(\kappa_i) = 0. 
\label{082}
\eeq
Let $\kappa_i\equiv 1$. This implies a simple relation between $\mathcal{H}_0$ and $\mathcal{R}_0$. Its solutions are $\mathcal{H}_0=1$ and $\mathcal{R}_0$ a free parameter; or
\beq
\mathcal{H}_0=\frac{1+2\mathcal{R}_0}{-1+2\mathcal{R}_0}.
\label{083}
\eeq
 If we take $\kappa_i= 1$ to be a critical point, then $\Pi'(1)=0$. It may be shown that when $\mathcal{H}_0=1$, $\Pi''(1)=0$.  Therefore, in order to exhibit unstable and stable solutions, we consider the set of solutions described by (\ref{083}). This leads to the two possibilities
\beq
\mathcal{R}^{\pm}_0 \equiv \frac{1}{2}(-3\pm 2\sqrt{2}).
\label{084}
\eeq
The solutions take the form
\beq
k=\cos \phi , \quad l=\sin \phi, \quad \xi =a \sin \phi , \quad \eta= a \cos \phi ,
\label{085}
\eeq
where $a$ is given by (\ref{073}). From (\ref{075}) we have $\phi= \phi_0 +\phi_1 t$ for $\phi_0$ and $\phi_1$ constants.
An example of this behavior is shown in figure 1, where the two potentials and corresponding solutions are shown. In one of these solutions the wave packet orbit lies inside the orbit of the vortex.  In the other solution, the opposite occurs.

\begin{figure} 
  \centering
  \includegraphics[width=2.5in]{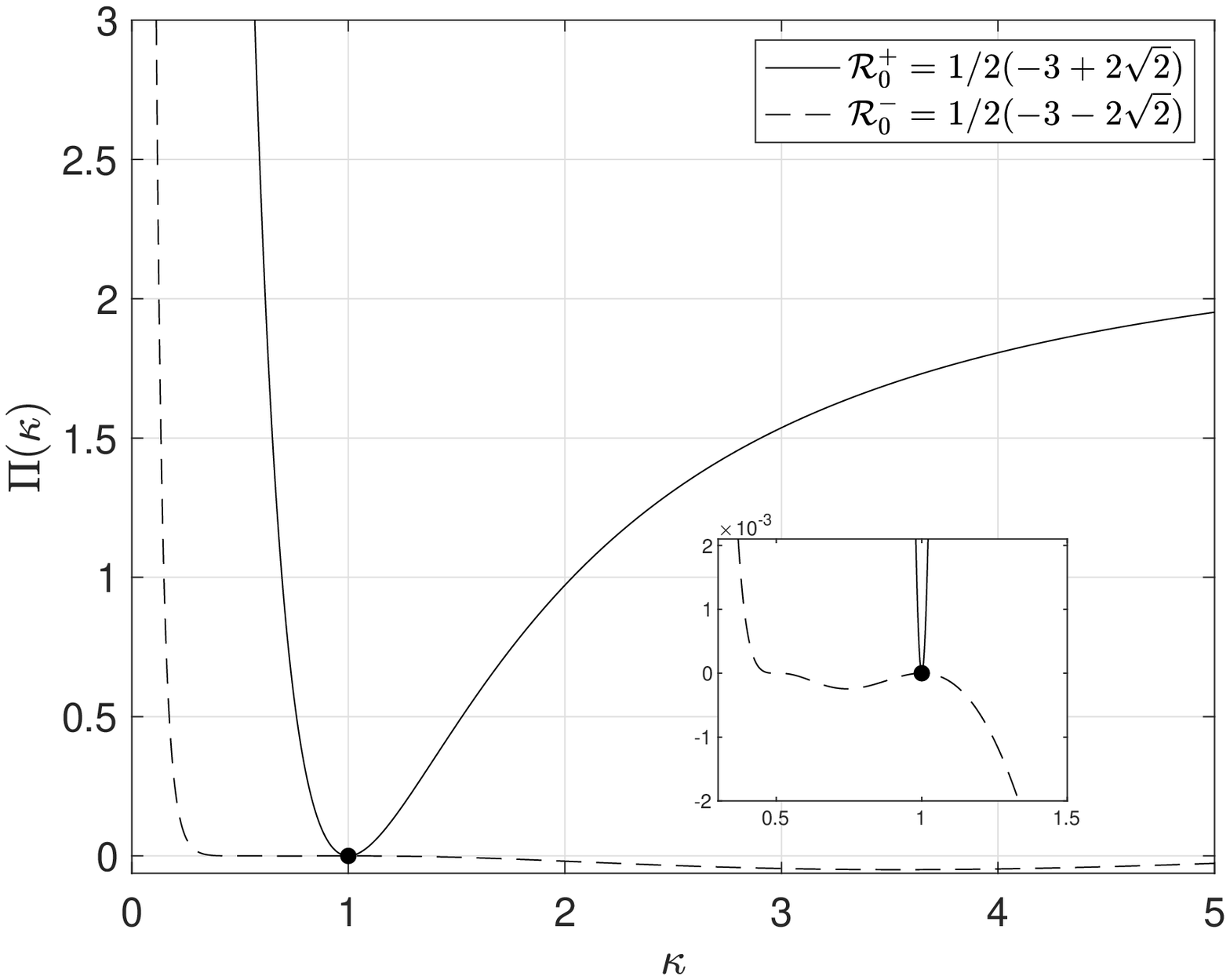}
 \includegraphics[width=2.5in]{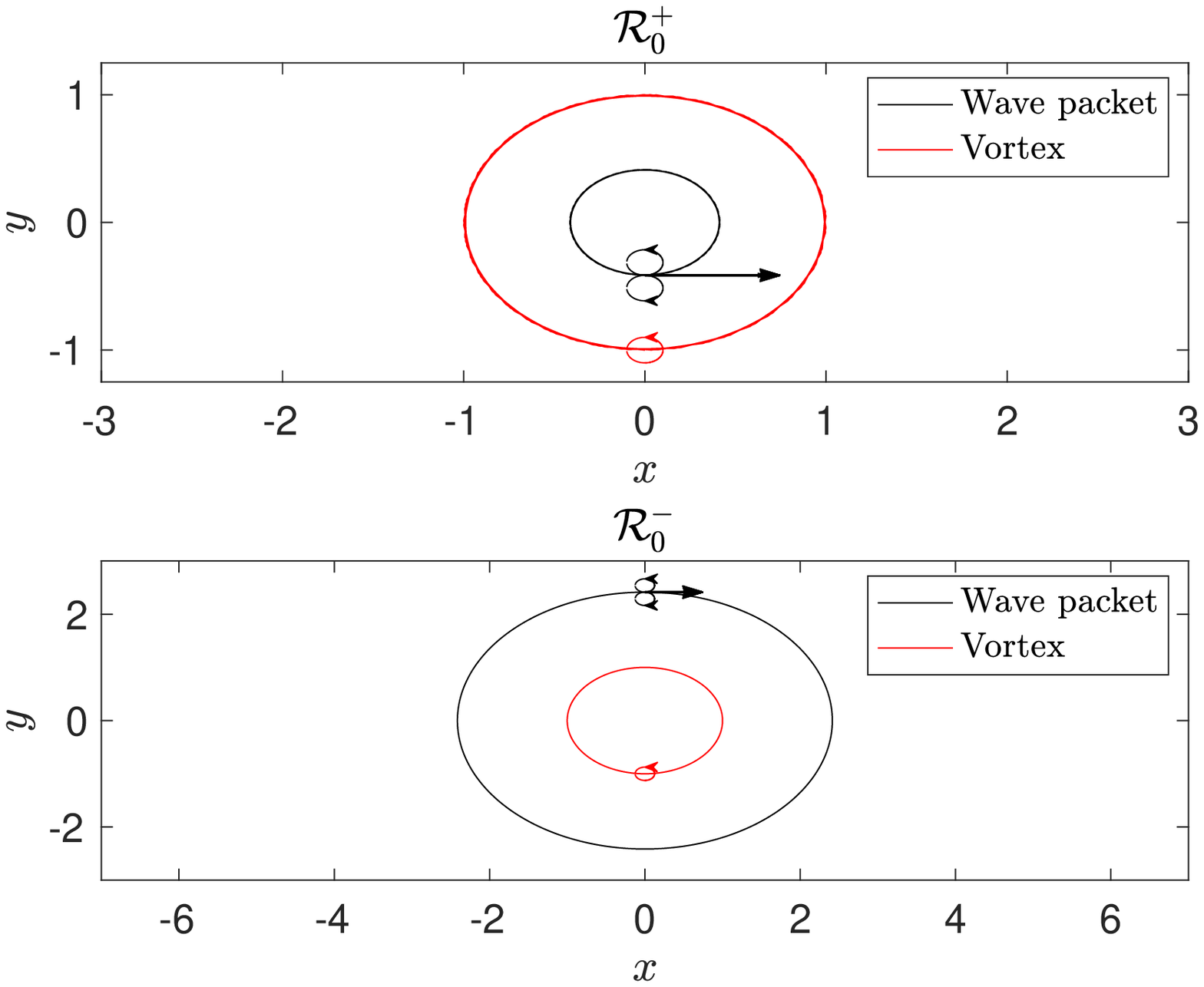}
  \caption{Left: The potentials $\Pi(\kappa)$ for the two cases ($\mathcal{R}^{\pm}_0$) in which a critical point is present at $\kappa=1$.  The inset enlarges the potentials near $\kappa=1$. In one case (dashed) the critical point coincides with a maximum of $\Pi(\kappa)$ (implying unstable motion) while in the other case (solid) the critical point is a minimum.  Right: Numerical solutions showing the locations of the wave packet (black line) and vortex (red line) in the two cases. The initial wave vector is indicated by the black arrow.  }
  \label{fig:1w1v}
\end{figure}

\subsection{Stability of orbits}
The $\Pi(\kappa)$ graphed in figure 1a suggest that the circular orbits shown there may not be locally stable (in a spectral sense) to perturbations. 
Therefore we study solutions in the neighborhood of $\kappa_0 = 1$. We take 
$\kappa =1 +\epsilon \kappa_1$ and expand $\Pi(\kappa)$ about the critical point $\kappa=1$, to find 
\beq
\Pi(\kappa) \approx \Pi(1)+\epsilon \kappa_1 \Pi'(1) + \epsilon^2 \kappa_1^2 \Pi''(1)+\cdots
\label{086}
\eeq
By \textcolor{black}{construction} $\Pi(1) = \Pi'(1) = 0 $. Taking $\kappa_1 = \kappa^0_1 e^{\lambda t}$, we find that the spectral stability of the system will be set by the sign of $\epsilon^2 (\kappa_1^0)^2 \Pi''(1)$. From figure 1 we see that $\mathcal{R}^+_0$ corresponds to stable orbits, and $\mathcal{R}^-_0$ corresponds to unstable orbits.  This is demonstrated in figure 2, which shows that the stable orbits are confined to the neighborhood of their initial trajectories, whereas the unstable orbits deviate considerably. 

\begin{figure} 
  \centering
  \includegraphics[width=2.5in]{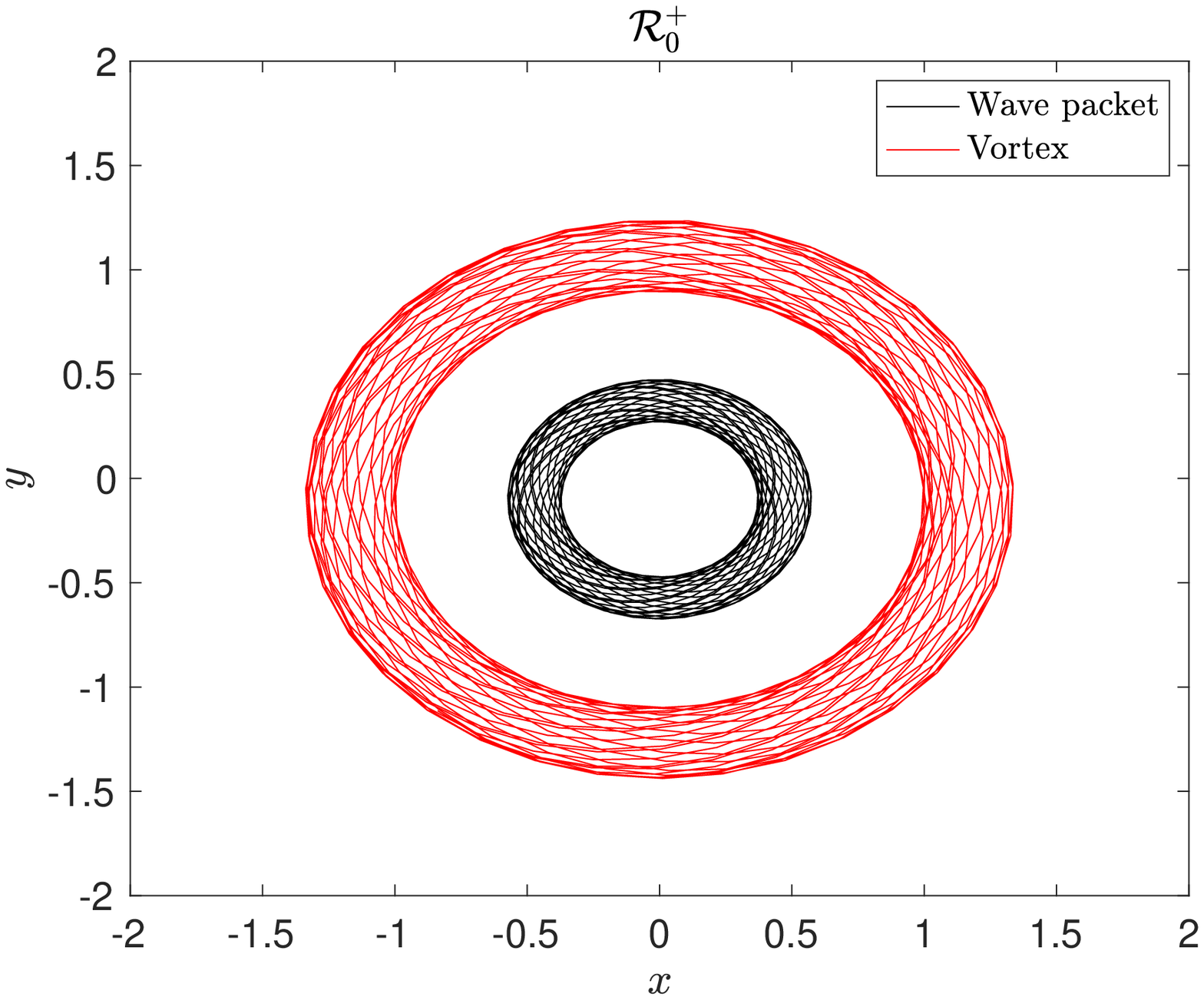}
 \includegraphics[width=2.5in]{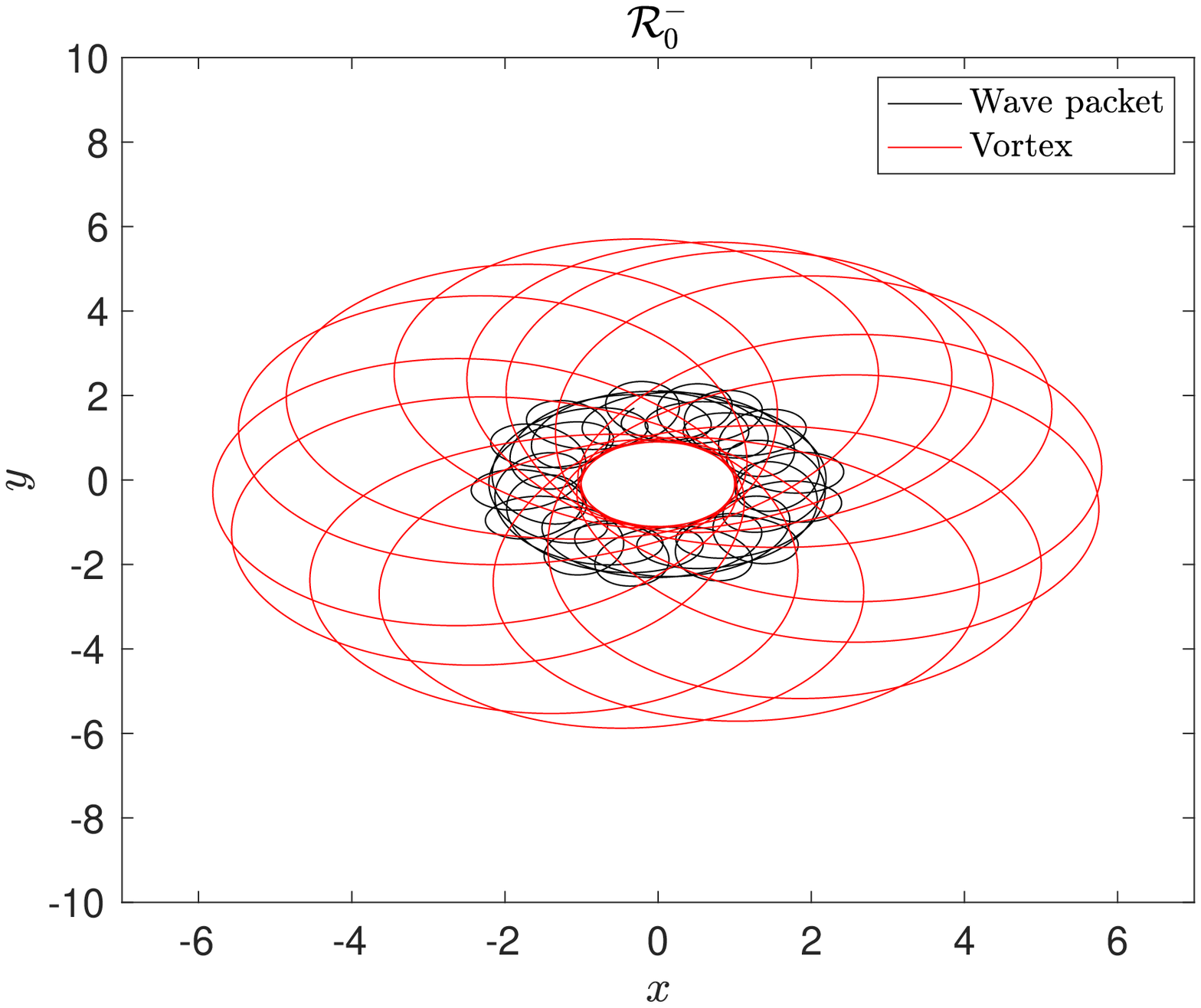}
  \caption{Left: Orbit of wave packet and vortex with $\mathcal{R}_0 = \mathcal{R}_0^+$ as defined in (\ref{085}). For this case perturbations are stable, and the orbits remain close to the unperturbed solution. Right: Orbit of wave packet and vortex with  $\mathcal{R}_0 = \mathcal{R}_0^-$. This orbit is unstable, and the solutions deviate considerably from circles.   }
  \label{fig:stability}
\end{figure}

\subsection{Collapse}

The above analysis addresses local spectral stability at a critical point. There are other solutions in which $a \to 0$ so that the wave packet and the vortex overlap. We call this phenomenon \textit{collapse}. Collapsed solutions violate the assumption of our model that the wave packets and vortices remain far apart.  Nonetheless, collapse is a real property of our equations that demands investigation. \textcolor{black}{Vortex collapse for three point vortices has been extensively studied \citep[see][and references therein]{Aref1983}}. The case of overlapping vorticity and wave action has been analyzed by \citet{Mcintyre2019}. 

The conditions for collapse are clear from equation (\ref{073}). Collapse occurs at the time $t^*$ at which
\beq
\kappa^2 \to 2\mathcal{R}_0.
\label{087}
\eeq
As an example we suppose that $\mathcal{H}_0=\mathcal{R}_0=0$.  Then the system collapses as $\kappa\to 0$. 
Under these assumptions, the governing equation for $\kappa$ reduces to 
\beq
\dot{\kappa}= \pm 2\sqrt{\frac{2-\sqrt{\kappa}}{\sqrt{\kappa}}}.
\label{088}
\eeq
Define $\vartheta$ by
\beq
\tan \vartheta = \pm \frac{\kappa^{1/4}}{\sqrt{2-\sqrt{\kappa}}}.
\label{089}
\eeq
Then  
\beq
\cos \vartheta = \pm \frac{1}{\sqrt{2}}\sqrt{2-\sqrt{\kappa}}, \quad \sin \vartheta =  \frac{1}{\sqrt{2}}\kappa^{1/4},
\label{090}
\eeq
and, solving for $\kappa$, we obtain
\beq
\kappa = \frac{1}{2}(3-4\cos 2\vartheta +\cos 4\vartheta).
\label{091}
\eeq
To find $t=t(\vartheta)$, we note that
\beq
\frac{dt}{d\vartheta}=\frac{dt}{d\kappa}\frac{d\kappa}{d\vartheta} = \pm \frac{1}{4}\tan \vartheta\left(\sin 2\vartheta +\frac{1}{2}\sin 4\vartheta \right).
\label{092}
\eeq
This can be integrated, and we arrive at
\beq
t=t_0 \pm \frac{1}{4}\left(12\vartheta -8\sin 2\vartheta + \sin 4\vartheta \right).
\label{093}
\eeq
Collapse occurs when 
\beq
 \frac{d \kappa}{d \vartheta}=\frac{dt}{d\vartheta} = 0.
 \label{094}
\eeq
Thus the collapse corresponds to the formation of a \textit{cusp} in $\kappa$. 

Figure 3 confirms these results. The top right panel shows the convergence of the wave packet and the vortex. The bottom right panel compares the theoretical prediction of $\kappa(t)$ (where we have taken the negative branch of the solution corresponding to $\dot{\kappa}<0$) to the numerical result.  The two curves are indistinguishable.  

In our model the wave action $\mathcal{A}$ is fixed.  Therefore, the wave energy $ \mathcal{A} \; \omega(\kappa)$ vanishes as $\kappa \to 0$ since $\omega(\kappa) \propto \sqrt{\kappa}$ for surface gravity waves.  The energy lost by the wave packet appears as an increase in the `interaction energy' between the wave packet and the vortex---an increase in the last term in \eqref{056}---but, again, the whole theory breaks down when the two particles finally converge. 

\begin{figure} 
  \centering
  \includegraphics[width=2.5in]{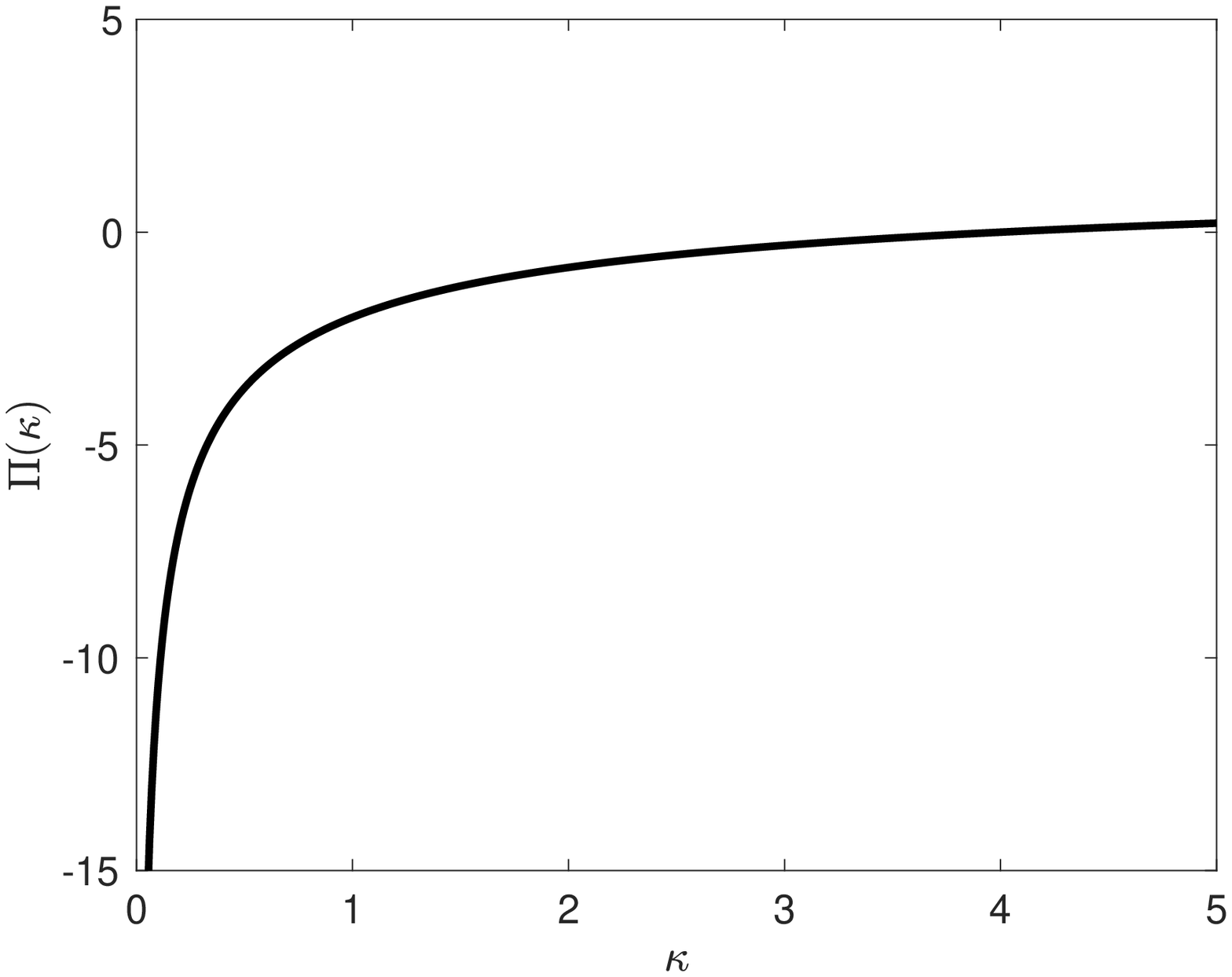}
 \includegraphics[width=2.5in]{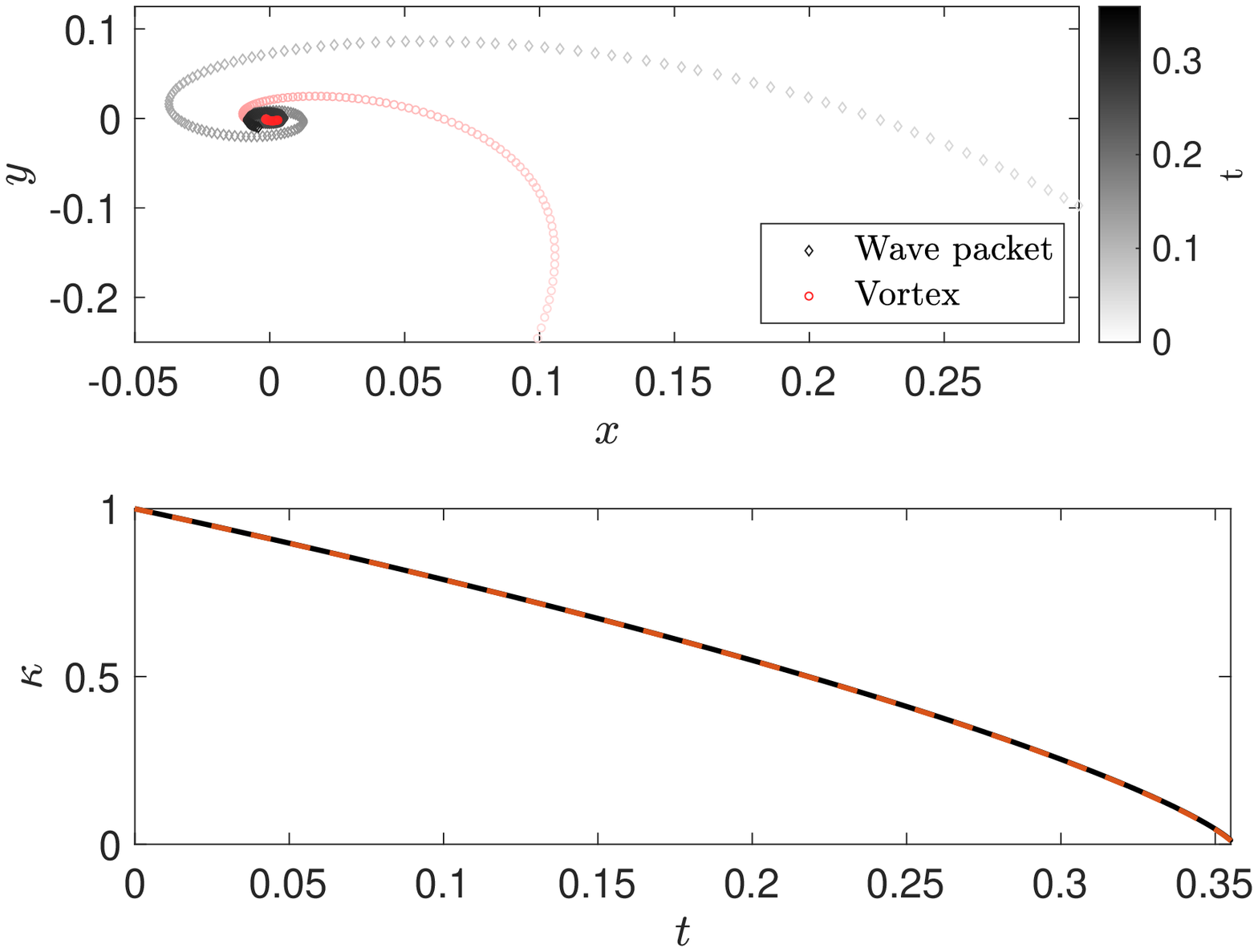}
  \caption{An example of ``collapse", in which the wave packet and the vortex converge, violating model assumptions.   Left: The potential $\Pi(\kappa)$ with $\mathcal{R}_0=0$ and $\mathcal{H}_0=0$. Note the singularity at $\kappa=0$.  The top right panel shows the converging particle paths. The bottom right panel shows the evolution of $\kappa(t)$, which vanishes in a cusp. The analytic solution is shown by the red dashed line and is indistinguishable from the numerical result. }
  \label{fig:stability}
\end{figure}

\subsection{Blow-up}

There are also solutions in which $\kappa$, the wave number modulus, grows without bound.  We call these \textit{blow-up} solutions. They correspond to wave packets that steepen. In reality, wave breaking limits the steepness of waves, and could be added to our model to extend its validity.  For example, wave packets that exceed a prescribed steepness could be replaced by counter-rotating vortices with a dipole moment determined by momentum conservation \eqref{052} as in \citet{Buhler2001}.  In this paper we consider only ideal solutions, and we do not include wave breaking.

A necessary condition for $\kappa(t)$ to grow without bound is that
\beq
\Pi'<0
\label{095}
\eeq
for large $\kappa$.
For large $\kappa$, 
\beq
\Pi \sim 2-\frac{4}{\sqrt{\kappa}}(1+2\mathcal{H}_0),
\label{096}
\eeq
hence blow-up requires
\beq
\mathcal{H}_0<-\frac{1}{2}.
\label{097}
\eeq
As an example, we take $\mathcal{H}_0=-1$.  Then
\beq
\Pi \sim 2+O(1/\sqrt{\kappa}),
\label{098}
\eeq
which implies that the blow-up solution takes the form
\beq
\kappa = \kappa_0 + \kappa_1 t .
\label{099}
\eeq
We examine this numerically in figure 4, and find agreement with the theoretical prediction. 

\begin{figure} 
  \centering
  \includegraphics[width=4.5in]{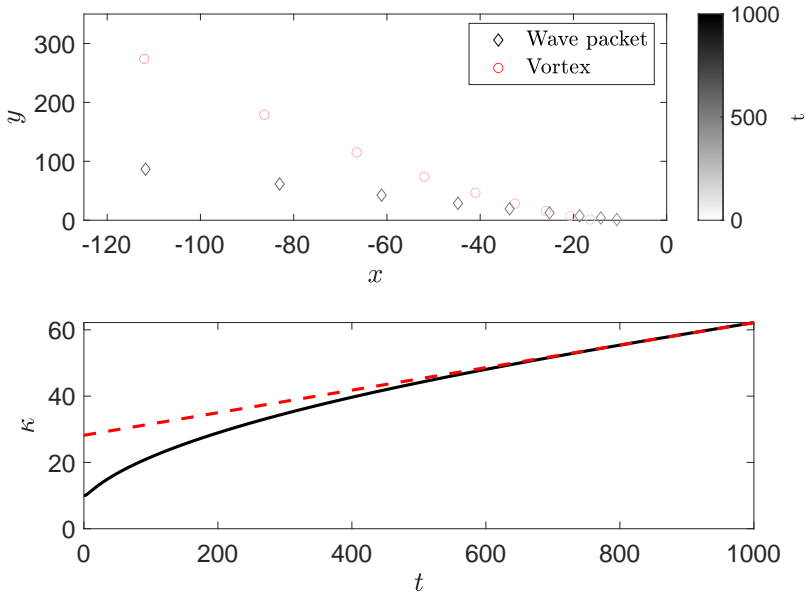}
  \caption{An example of a solution that ``blows up", meaning $\kappa \to \infty.$ The trajectories of the wave packet and vortex are shown in the top panel. In the bottom panel, we display $\kappa$. The asymptotic form of the growth is predicted to go like $t$, which is shown by the dashed red line and is seen to agree well with the numerical integration.}
  \label{fig:stability}
\end{figure}

\section{Two vortices and one wave packet}

We now examine the system comprising a single wave packet with action $\mathcal{A}_p$ and wavevector $(k_p, l_p)$ located at $(x_p, y_p)$; a point vortex of strength $-\Gamma$ located at $x_1,y_1$; and a second point vortex of strength $+\Gamma$ located at $x_2,y_2$.  Refer to figure 5.
Initially,
\beq
y_p=l_p=0, \;\;\; x_2=x_1, \;\;\;\; y_2=-y_1
\label{100}
\eeq
and, by symmetry, these conditions hold at all later times.  
The Lagrangian is \textcolor{black}{(with $H_0=1$)}
\begin{align}
&L[x_p,y_p,k_p,l_p,x_1,y_1,x_2,y_2]=
\notag \\
&\int dt  \; \Big[
\mathcal{A}_p \left( k_p \dot{x}_p+l_p \dot{y}_p -\omega_r(k_p,l_p) \right) 
+\Gamma \left( {x}_1\dot{y}_1-{x}_2\dot{y}_2 \right)
-\frac{\Gamma^2}{2\pi} \ln\vert \mathbf{x}_1-\mathbf{x}_2 \vert
\notag \\
&  \;\;\;\;\;\;\;\;    +\frac{\mathcal{A}_p}{2\pi} \left( 
-\Gamma \frac{(\mathbf{x}_1-\mathbf{x}_p) \times \mathbf{k}_p} {\vert \mathbf{x}_1-\mathbf{x}_p \vert ^2}
+\Gamma \frac{(\mathbf{x}_2-\mathbf{x}_p) \times \mathbf{k}_p} {\vert \mathbf{x}_2-\mathbf{x}_p \vert ^2}
\right) \Big].
\label{101}
\end{align}
We vary all the dependent variables, and \emph{then} apply the symmetry conditions \eqref{100} to obtain a closed set of four equations.  (It is illegal to apply the symmetry condition before taking the variations.)  The reduced set of equations is
\begin{align}
&\delta k_p: \;\;\;\;   \dot{x}_p= c_g(k_p,0)-\frac{\Gamma y_1}{\pi d^2},
\label{102} \\
&\delta x_p: \;\;\;\;   \dot{k}_p=\frac{2\Gamma k_p}{\pi} \left(
\frac{(x_1-x_p)y_1}{d^4}  \right),
\label{103} \\
&\delta x_1: \;\;\;\;   \dot{y}_1=\frac{\mathcal{A}_p k_p}{\pi}
 \frac{(x_1-x_p)y_1}{d^4},
\label{104} \\
&\delta y_1: \;\;\;\;   \dot{x}_1=-\frac{\Gamma}{4\pi y_1}
+\frac{\mathcal{A}_p k_p}{2\pi} \frac{(x_1-x_p)^2-y_1^2}{d^4},
\label{105}
\end{align}
where $c_g$ is the $x$-component of the group velocity, and
\beq
d^2 \equiv (x_1-x_p)^2+y_1^2
\label{106}
\eeq
is the squared distance between the wave packet and either vortex.
 Because of the symmetry conditions \eqref{100}, we do not need the evolution equations for $y_p, l_p, x_2,$ and $ y_2$.

The equations \eqref{102}-\eqref{105} conserve energy in the form
\beq
E=\omega_r(k_p) \mathcal{A}_p+\frac{\Gamma^2}{2\pi} \ln y_1 
-\frac{\mathcal{A}_p\Gamma}{\pi} \frac{y_1 k_p}{d^2}
\label{107}
\eeq
and momentum in the form
\beq
\mathcal{M}=\mathcal{A}_p k_p -2\Gamma y_1.
\label{108}
\eeq
The angular momentum vanishes.
Defining
\beq
X(t)\equiv x_p(t)-x_1(t),
\label{109}
\eeq
we rewrite \eqref{102}-\eqref{105} as three equations
\beq
 \dot{X}= c_g(k_p)+\frac{\Gamma}{4\pi d^2 y_1}(X^2-3y_1^2)
-\frac{\mathcal{A}_p k_p}{2\pi d^4} (X^2-y_1^2),
\label{110} 
\eeq
\beq
\dot{k}_p=-\frac{2\Gamma k_p}{\pi d^4}Xy_1,
\label{111} 
\eeq
\beq
 \dot{y}_1=-\frac{\mathcal{A}_p k_p}{\pi d^4} Xy_1,
\label{112}
\eeq
in the three unknowns $k_p$, $y_1$ and $X$, where now $d^2= X^2+y_1^2$. 
The two conserved quantities, \eqref{107} and \eqref{108}, make this an integrable system.  Eliminating $y_1$ between \eqref{107} and \eqref{108}, we obtain an expression for the energy in terms of $X$ and $k_p$.  The motion is confined to curves of constant $E(X,k_p)$.  We can determine the solution by considering $E(X,k_p)$ or, even more conveniently, by considering
\beq
E(k_p,d^2)=\omega_r(k_p) \mathcal{A}_p
+\frac{\Gamma^2}{2\pi} \ln (\mathcal{A}_pk_p -\mathcal{M}) 
-\frac{\mathcal{A}_pk_p}{2\pi} \frac{(\mathcal{A}_pk_p-\mathcal{M})}{d^2},
\label{113}
\eeq
in which we have dropped additive constants.  Only the last term in \eqref{113} involves $d^2$.

Consider a gravity wave packet, initially at $X=-\infty$ with $k_p>0$, approaching the vortex pair from the left, as shown in the top panel of figure 5.
While the wave packet is still far from the vortex pair ($d^2$ very large) the last term in \eqref{113} is negligible.  According to \eqref{112}, $k_p$ increases with time on $X<0$.  This increase in $k_p$ occurs because the velocity field associated with the vortices squeezes the wave packet in the $x$-direction.  Since $c_g (k_p)>0$ the wave energy $\omega_r \mathcal{A}_p$ and the vortex-interaction energy---the middle term in \eqref{113}---both increase with $k_p$.  The increase in the latter corresponds to the two vortices being pushed apart by the velocity field associated with the dipole.  The increase in these two terms must be balanced by the last term in \eqref{113}, which represents the energy stored in the superposed velocity fields of the wave packets and vortices.  These superposed fields tend to cancel as the wave packet approaches the vortex pair.  $k_p$ reaches its maximum value at $X=0$, where 
\beq
d^2=y_1^2=\left( \frac{\mathcal{A}_pk_p -M}{2\Gamma} \right)^2.
\label{114}
\eeq
Substituting \eqref{114} into \eqref{113}, we obtain an equation for this maximum value of $k_p$.  After passing $X=0$, the solution `unwinds', and $k_p$ returns to its original value as $X \rightarrow \infty$.
The numerical solution shown in figure 5 confirms this analysis.

\begin{figure} 
  \centering
  \includegraphics[width=4.5in]{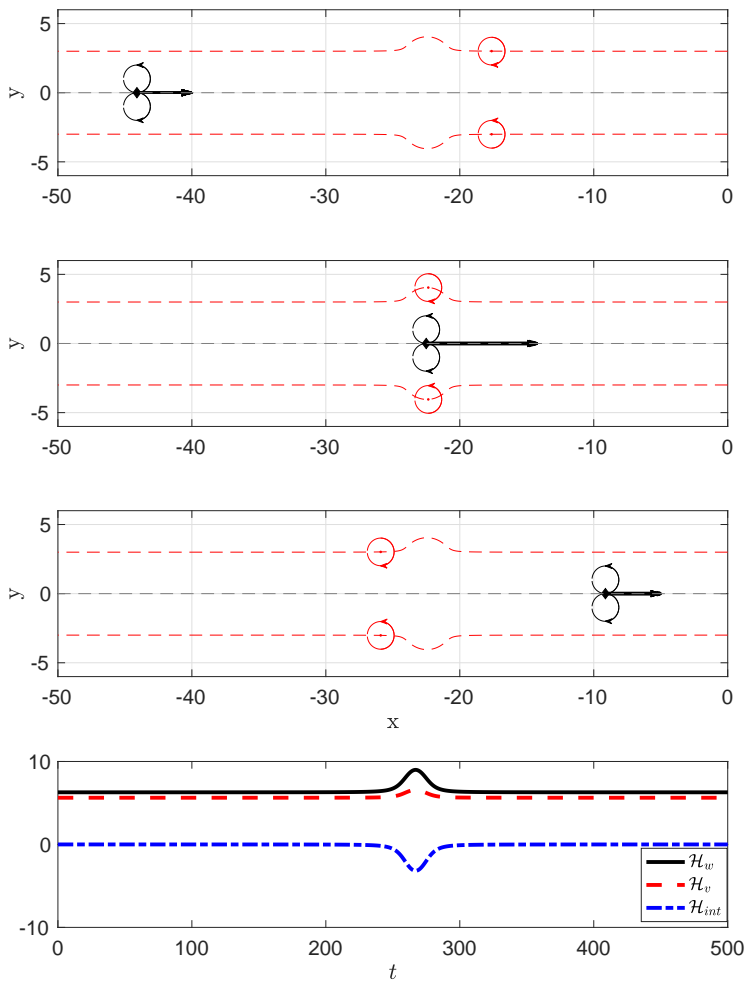}
  \caption{Top panel: A right-moving wavepacket (in black), with its wave vector denoted by the straight arrow, and its associated dipolar flow indicated by circles, collides with a left-moving pair of counter-rotating vortices (in red). As the wave packet approaches the vortices, the flow induced by the vortex pair squeezes the wave packet in the $x$-direction, stretching its wave vector. The dipolar flow induced by the wave packet pushes the vortices apart (second panel). After passage of the wave packet (third panel) the solution `unwinds', and all three particles return to their original configurations.  The bottom panel partitions the energy \eqref{114} into wave energy $\mathcal\mathcal{H}_w$ (the first term on the left hand side of \eqref{114}), vortex energy $\mathcal{H}_v$ (the second term), and interaction energy $\mathcal{H}_{int}$. $\mathcal\mathcal{H}_w$ and $\mathcal{H}_v$ increase during the interaction, while $\mathcal{H}_{int}$  decreases. }
  \label{fig:stability}
\end{figure}

\subsection{Wave-packet induced drift }
 
If $\Gamma \ll \textcolor{black}{\mathcal{A}}$,  (\ref{108}-\ref{112}) imply 
\beq
\dot{X} = c_g(k_p) -\frac{\mathcal{A}_pk_p}{2\pi d^4} (X^2-y_1^2),
\label{116}
\eeq
\beq
\dot{k}_p = 0,
\label{117}
\eeq
\beq
\dot{y}_1 = -\frac{\mathcal{A}_p k_p}{\pi d^4}Xy_1.
\label{118}
\eeq
Hence $k_p$ and $c_g$ are constants. The  governing equations (with $\mathcal{A}_p=2\pi a^2$ and $k_p = g = 1$) reduce to
\beq
\dot{X} = c_{g0} -\frac{a^2}{d^4} (X^2-y_1^2),
\label{119}
\eeq
\beq
\dot{y}_1 = -\frac{2 a^2}{d^4}Xy_1.
\label{120}
\eeq

In this limit, the point vortices are passive; their motion is the same as that of fluid particles in the presence of a uniformly translating cylinder.  This problem was examined by 
\citet[see also \citealt{Morton1913, Darwin1953}]{Maxwell1870}.
In the reference frame moving with the wave packet, the stream function is an integral of motion. Hence 
\beq
Y_0=y_1\left(1-\frac{a^2}{d^2}\right)
\label{121}
\eeq
is  constant. Define $\dot{\mathcal{X}} = \dot{X}-c_{g0}$ and  $\theta = \tan ^{-1}(y_1/X)$.  Using \eqref{121} \textcolor{black}{and following \citet{Maxwell1870} and \citet{Darwin1953}} we find that
\beq
\mathcal{X} = \int \frac{a^2\cos 2\theta }{\sqrt{Y_0^2+4a^2\sin^2\theta}}\ \dd \theta.
\label{122}
\eeq
Let $\cos \theta =- \text{sn} (\nu)$ with the suppressed modulus of the Jacobi elliptic function understood to be
\beq
m=\kappa^2=\frac{4a^2}{Y_0^2+4a^2}.
\label{123}
\eeq
It is tedious but straight-forward to show that
\beq
\mathcal{X}=\frac{a}{\kappa}\left(\left(1-\frac{\kappa^2}{2}\right) \nu -E(\nu)\right).
\label{124}
\eeq
Similar expressions may be found for $y_1(\nu)$ and $t(\nu)$ but fall outside the scope of our discussion \citep{Darwin1953}. 

From (\ref{124}), the total drift in the $x$-direction is
\beq
\Delta \mathcal{X} = \frac{2a}{\kappa} \left(\left(1-\frac{\kappa^2}{2}\right) K-E\right),
\label{127}
\eeq
where $E,K$ are the complete elliptical integrals of the first and second kind, respectively. 

The drift volume, $D$, defined as
\beq
D=\int_{-\infty}^{\infty}\mathcal{X} \ \dd Y = \pi a^2, 
\label{128}
\eeq
has the same order of magnitude as the vertically integrated Stokes drift. Note, the connection between Stokes drift \textcolor{black}{(specifically the motion in the \textit{vertical} plane)} and Darwin drift has been examined by \citet{Eames1999}. 

\section{1 vortex, N wave packets}

We now search for simple harmonic motion in a system comprising one vortex and $N>1$ wave packets \textcolor{black}{\citep[see also the related discussion on a ring of geostrophic vortices in][]{Morikawa1971}}. The single vortex of strength $\Gamma=  2\pi$ remains stationary at $\bold{x}=\bold{0}$. The $N$ wave packets at $\bold{x}_p=(x_p,y_p)$ have equal actions $\mathcal{A}_p=\mathcal{A}$, and wave vectors of equal magnitude $|\bold{k}_p|=\kappa$. They lie symmetrically on the circle $|\bold{x}_p|=\chi$. Both $\kappa$ and $\chi$ are constants.  We take the arbitrary depth parameter to be unity, $H_0=1$. 

A vortex at $\bold{x}$ induces the velocity field
\beq
\bold{U}_m(\bold{x}', \bold{x} )=
\frac{(y-y',x'-x)}{(x-x')^2+(y-y')^2}
\label{129a}
\eeq
at $\bold{x}'$.
Hence
\beq
\bold{U}_m(\bold{x}_p, \bold{0} ) = \frac{(-y_p,x_p)}{\chi^2}.
\label{129b}
\eeq
We also need
\beq
\nabla U_m(\bold{x}_p, \bold{0} ) =  \frac{1}{\chi^4}
\left( 2x_p y_p , x_p^2- y_p^2 \right)
\eeq
and
\beq
\nabla V_m(\bold{x}_p, \bold{0} ) =  \frac{1}{\chi^4}
\left( -x_p^2+ y_p^2 , -2x_p y_p \right).
\eeq
The wavepacket at $\bold{x}_p$ induces the velocity 
\beq
\bold{U}_d 
=- \frac{\mathcal{A}}{2\pi\chi^4}
\left( 2l_px_py_p+k_p(x_p^2-y_p^2) \; , \; 2k_px_py_p-l_p(x_p^2-y_p^2) \right)
\label{131}
\eeq
at the vortex.

The equations of motion reduce to 
\beq
\dot{\bold{x}}_p=\frac{1}{2}\sqrt{\frac{g}{\kappa^3}}(k_p,l_p)+\frac{(-y_p,x_p)}{\chi^2},
\label{133}
\eeq
\beq
\chi^4\dot{\bold{k}}_p=(-2k_px_py_p+l_p(x_p^2-y_p^2),2l_px_py_p+k_p(x_p^2-y_p^2)),
\label{133}
\eeq
and
\beq
\sum_p \mathcal{A}(-2l_px_py_p-k_p(x_p^2-y_p^2),-2k_px_py_p+l_p(x_p^2-y_p^2))=\bold{0}.
\label{134}
\eeq
The last equation is the condition that the vortex remains stationary. 

We look for solutions exhibiting simple harmonic motion with $\bold{k}_p=\kappa (\sin \theta_p,-\cos \theta_p)$ and $\bold{x}_p=\chi(\cos \theta_p, \sin \theta_p)$.
This leads to the constraints  
\beq
\kappa =  \frac{\chi^2}{16},
\label{135}
\eeq
\beq
\frac{d\theta_{p}}{dt}= - 1/\chi^2,
\label{136}
\eeq
and
\beq
\sum \mathcal{A} (\cos \theta_p,\sin \theta_p) = \bold{0}.
\label{137}
\eeq
Taking $\mathcal{A}=1$, we observe that solutions to this system are related to the $N$-th roots of unity. This sets the initial phase of $\theta_p$. 
We find that
\beq
\mathbfcal{M}=\bold{0}; \quad \mathcal{H} = \frac{3\chi }{16}\sum \frac{\mathcal{A}_p}{2\pi}. 
\label{138}
\eeq
Figure 6 shows a solution of this type with four wave packets.

\begin{figure} 
  \centering
  \includegraphics[width=3.in]{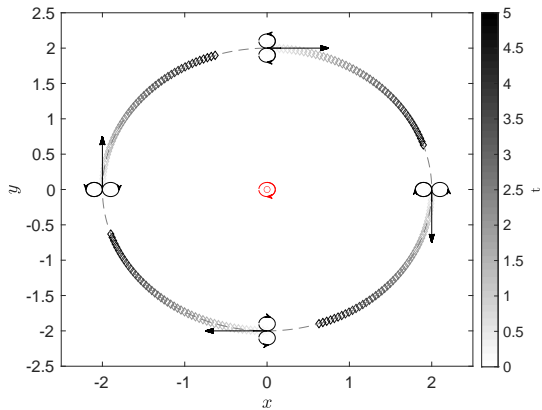}
  \caption{Trajectories exhibiting circular motion for a system with four wave packets and one vortex. The wave packets are shown in black, with their intensity increasing with time. The red circle represents the stationary vortex.  }
  \label{fig:stability}
\end{figure}

\section{The Lagrangian motion of a particle near a periodic wave packet}

We now consider the motion of a very weak point vortex near a periodic array of wave packets. In the limit of vanishing circulation, the vortex acts as a passive tracer, and sets the foundation for the stability analysis performed in \S 7. This is a generalization of the motion considered by \citet{Maxwell1870} and \citet{Darwin1953}. Unlike the analysis there, closed form solutions for the motion of a particle are not found, but asymptotic analysis reveals interesting features of the induced flow. 

The system we now consider is unbounded in $x$ and $y$, and infinitely periodic in the  $x$-direction. Initially, the wave packet propagates along the $x$-axis. Each vortex or wave packet at $(x,y)$ sees its images at $(x+2\pi n,y)$, where $n$ is any integer. Again we take $H_0=1$. 

The governing equations are (\ref{047}) - (\ref{049}) with $\bold{U}_m$ and $\bold{U}_d$ now calculated from the stream functions
\beq
\psi_m(\bold{x},\bold{x}_i) = \frac{1}{4\pi} \sum_n \ln [(x-x_i+2\pi n)^2 + (y-y_i)^2],
\label{141}
\eeq
and
\beq
\psi_d (\bold{x},\bold{x}_p,\bold{k}_p) = \frac{1}{2\pi}\sum_n  \frac{(\bold{x}-\bold{x}_p+2\pi(n,0))\times \bold{k}_p}{|\bold{x}_p-\bold{x}+2\pi(n,0)|^2}.
\label{142}
\eeq

\subsection{Limit of weak point vortices}
We begin by considering the wave-packet-induced motion of the vortices when 
$|\Gamma_i| \ll 1$. As in \S 4---see particularly \S 4.1---this motion takes a nontrivial form. 

We consider a wave packet traveling along the x-axis so that $l=y_p=0$. To $O(1)$, (\ref{047})-(\ref{049}) reduce to 
\beq
\dot{x}_p = \frac{1}{2}\sqrt{\frac{g}{|\bold{k}_p|}} \frac{\bold{k}_p}{|\bold{k}_p},
\label{146}
\eeq
\beq
\dot{\bold{k}}_p=0,
\label{147}
\eeq
\beq
\dot{\bold{x}}_i = \bold{U}_d(\bold{x}_i, \bold{x}_p, \bold{k}_p   ),
\label{148}
\eeq
where $\bold{U}_d$ is computed from \eqref{142}. As we are considering one (weak) vortex per period, $i=1$ and we take $y_1=y$.

It is convenient to define 
\beq
\chi = x_1 -x_p.
\label{147}
\eeq
The governing equations become 
\beq
\dot{\chi} = -c_{g0}+\mu \sum_{n=-\infty}^{\infty} \frac{(\chi-2\pi n)^2-y^2}{((\chi-2\pi n)^2+y^2)^2},
\label{148a}
\eeq
\beq
\dot{y} = 2\mu \sum_{n=-\infty}^{\infty}  \frac{(\chi-2\pi n)y}{((\chi-2\pi n)^2+y^2)^2},
\label{149b}
\eeq
Defining $z = \chi+\ii y$, we have 
\beq
\dot{z} = -c_{g0}+\mu\sum_{n=-\infty}^{\infty} \frac{(z-2\pi n)^2}{(z-2\pi n)^2(z^*-2\pi n)^2}=-c_{g0}+\mu\sum_{n=-\infty}^{\infty} \frac{1}{(z^*-2\pi n)^2}.
\eeq
Define the complex-valued velocity potential $w=\phi +\ii \psi$ where $(w_z)^*\equiv \dot{z}$. This implies \citep[\S 64]{Lamb1932}
\beq
w=-c_{g0}\chi -\mu \sum_{n=-\infty}^{\infty}\frac{1}{2\pi n-z} = -c_{g0}\chi - \mu \cot\frac{z}{2}.
\label{153}
\eeq
Eqn \eqref{153} implies 
\beq
\phi =- c_{g0}\chi + \mu\frac{\sin \chi }{ \cos \chi - \cosh y},
\label{154}
\eeq
and
\beq
\psi = - c_{g0}y- \mu \frac{\sinh y }{\cos \chi - \cosh y}.
\label{155}
\eeq
From the relation $(w_z)^*=\dot{z}$, we find
\beq
\dot{\chi} =-c_{g0}+ \frac{\mu}{2} \frac{1-\cos \chi \cosh y}{(\cos \chi - \cosh y)^2},
\label{148}
\eeq
\beq
\dot{y} =\frac{\mu}{2} \frac{\sin \chi \sinh y}{(\cos \chi - \cosh y)^2}.
\label{149}
\eeq

The stream function is a material contour. As in \S 4.1, this provides an additional conserved quantity. When  $(\chi_i,y_i)$ are small---vortex $i$ very close to the wave packet---our equations reduce to the equations of Maxwell discussed in \S4.1. 

Although we have been unable to find closed form solutions, asymptotic analysis reveals interesting properties of this system. We expand the transcendental pieces of \eqref{148} and \eqref{149} as 
\beq
\frac{\sinh y\sin \chi}{(\cosh y - \cos \chi)^2} = 2  \sum_{n=1}^{\infty} ne^{-ny}\sin n\chi
\label{156}
\eeq
and
\beq
\frac{1-\cosh y\cos \chi}{(\cosh y - \cos \chi)^2} =-2 \sum_{n=1}^{\infty} ne^{-n y}\cos n\chi.
\label{157}
\eeq
This expansion assumes $y > 0$ ; a similar formula holds for $y<0$.
The equations of motion may then be written as 
\beq
\dot{\chi}=-c_{g0}-\mu  \sum_{n=1}^{\infty} n e^{-ny}\cos n\chi, 
\label{158}
\eeq
\beq
\dot{y}= \mu  \sum_{n=1}^{\infty} n e^{-ny}\sin n\chi,
\label{159}
\eeq
or, more compactly, as
\beq
\dot{z} = -c_{g0}-\mu \sum_{n=1}^{\infty} n e^{- \ii n z^*}.
\label{160}
\eeq
Let $b>0$ to be the initial vertical coordinate of the particle, and  (for reasons that become clear in \S 7) let $\pi/2$ be its horizontal coordinate. Let the initial location of the wave packet be $(\pi,0)$. 

As the wave packets and vortex are sufficiently far apart, a natural small parameter arises, namely
\beq
\epsilon \equiv e^{-b}\ll 1.
\label{161}
\eeq
We expand $z$ as
\beq
z = z_{0}+\epsilon z_{1}+\epsilon^2 z_{2}+\cdots
\label{162}
\eeq
To $O(\epsilon^2)$, we find 
\beq
\dot{z} =\dot{z}_{0}+\epsilon \dot{z}_{1}+\epsilon^2 \dot{z}_{2}=  -c_{g0} - \mu \left(\epsilon e^{-\ii \chi_0} +\epsilon^2(2e^{-2\ii \chi_0}-\ii z_{1}^*e^{-\ii \chi_0}) \right).
\label{162}
\eeq
We solve this system iteratively. To lowest order,
\beq
z_0 = z_0^0-c_{g0}t,
\label{163}
\eeq
where $z_0^0=-\pi/2+ \ii b $. This implies
\beq
z_1 =- \frac{\ii \mu}{c_{g0}}\left(e^{-\ii \chi_0}-\ii\right) =-\frac{\mu}{c_{g0}}(e^{\ii \theta}-1),
\label{164}
\eeq
where $\chi_0 = -\pi/2-c_{g0}t$ and we define
\beq
\theta \equiv c_{g0}t.
\label{165}
\eeq
Thus  $z_1$ takes the form of a simple harmonic oscillator. The constant is taken to ensure that $z_1=0$ at $t=0$.

At second order we find
\beq
\dot{z}_2 = \mu (2 e^{2\ii \theta} - z_1^* e^{\ii \theta}), 
\label{166}
\eeq
so that upon substitution of the result for $z_1$, we find
\beq
z_2 = \frac{\mu^2}{c_{g0}}t-\ii \frac{\mu}{ c_{g0}}\left(e^{2\ii\theta  }-1\right) + \ii \frac{\mu^2}{ c_{g0}^2} \left(e^{\ii \theta}-1\right).
\label{172}
\eeq
At this order there is a mean drift in the direction of wave propagation. Unlike the Stokes drift which is $O(\mathcal{A})$, this term is $O(\mathcal{A}^2)$. This has potentially important implications for the stability of vortex streets, as discussed in \S 7.

Figure 7 compares our approximate analytical solutions to numerical integrations of the full equations for  $\epsilon = 1/10$ (left panel) and $\epsilon =1/3$ (right panel). We see that the asymptotic theory works relatively well for small values of $\epsilon$. 

\begin{figure} 
  \centering
  \includegraphics[width=2.55in]{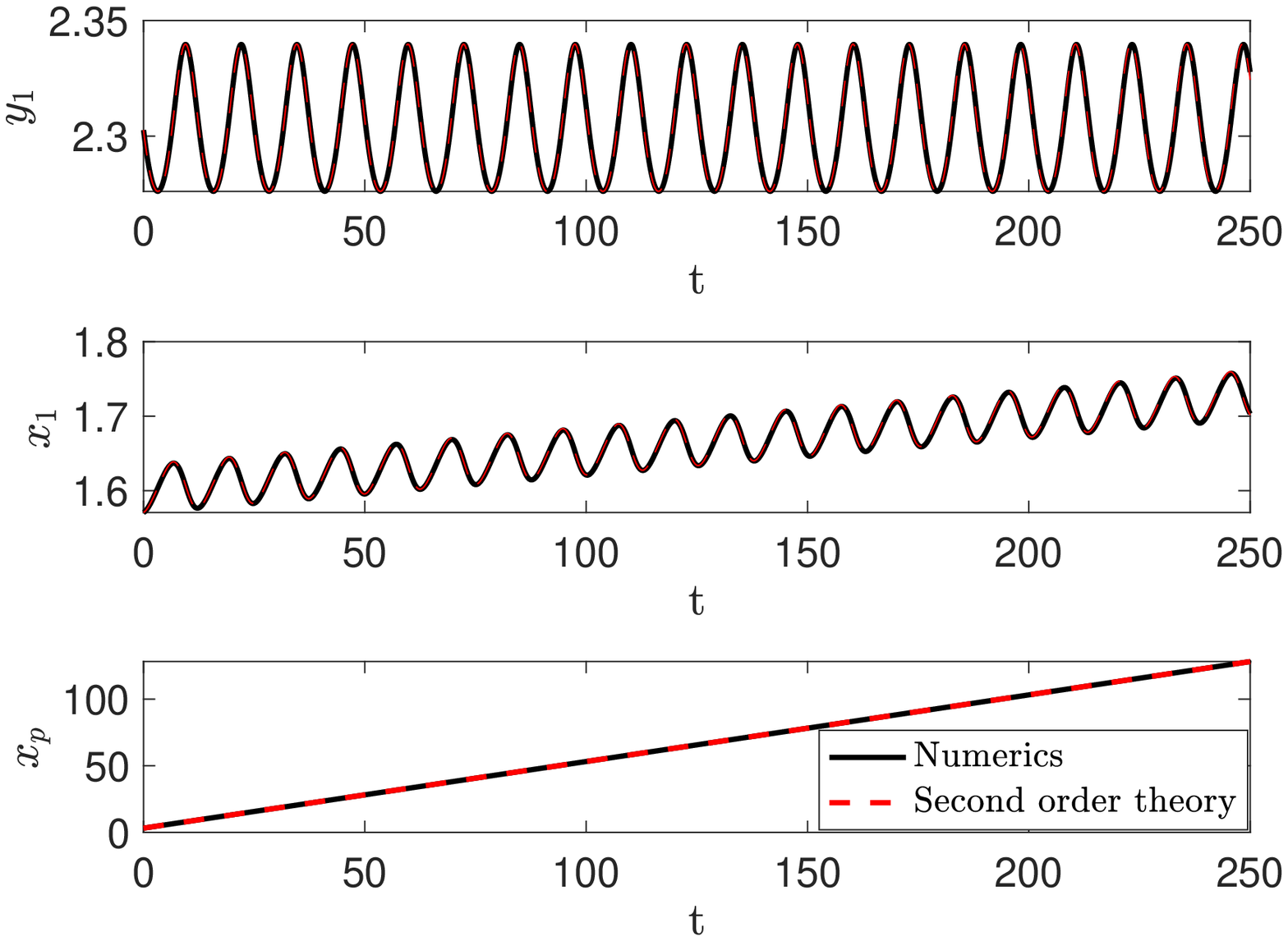}
    \includegraphics[width=2.55in]{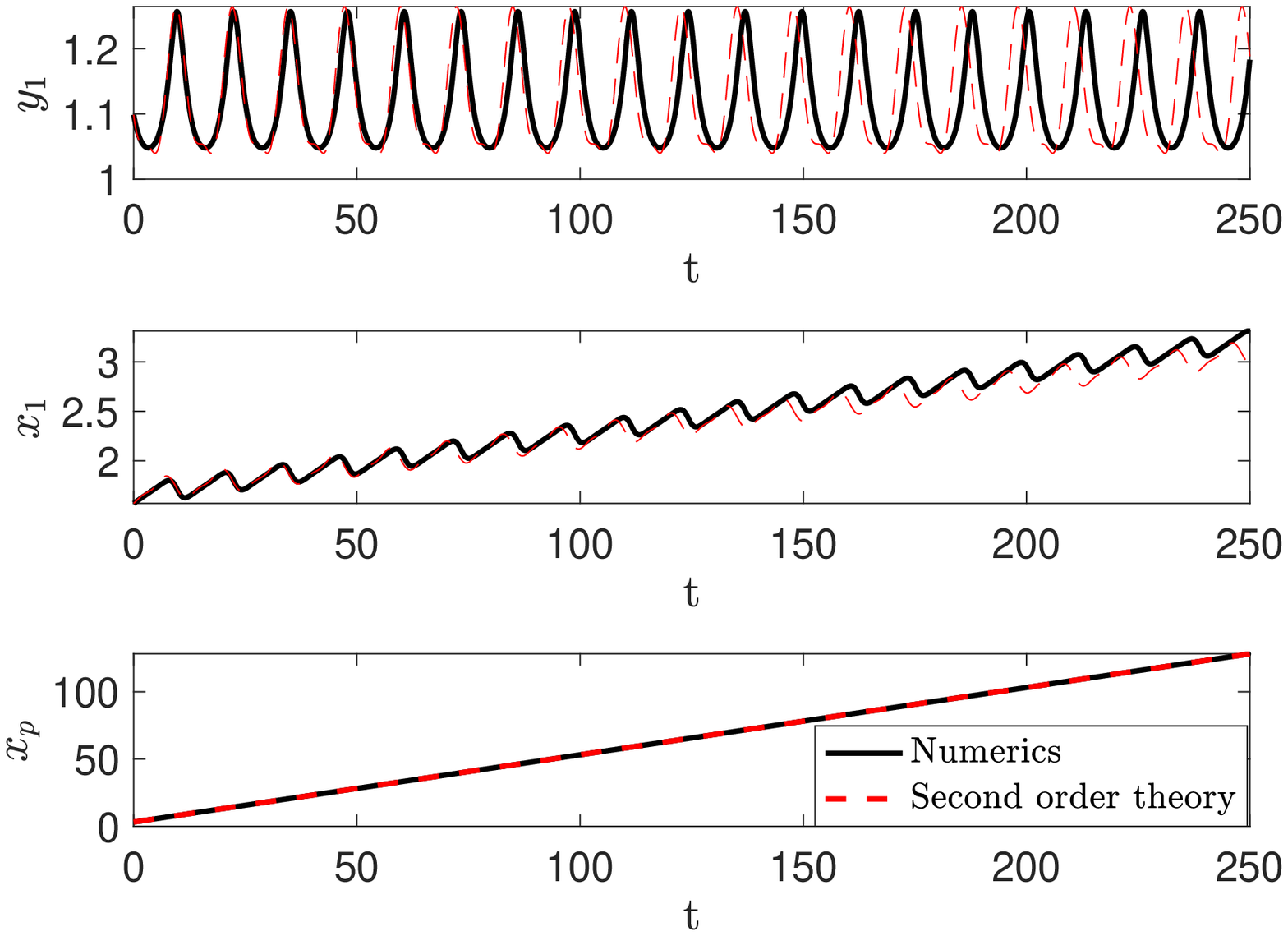}
  \caption{A comparison of the numerical integration of the full equations of motion (black lines) and the second order asymptotic solutions (dashed-red lines). The left column shows the results when $\epsilon =1/10$, while the right column shows the results for $\epsilon =1/3$. The top two row shows $y_1$ and $x_1$, the vertical and horizontal motion of the vortex, as a function of time. The bottom row shows the behavior of the wave packet. We see that the asymptotic theory describes the numerical results well for the case of $\epsilon =1/10$ but begins to break down for $\epsilon = 1/3$.  }
  \label{fig:stability}
\end{figure}

\section{The stability of a symmetric vortex street in the presence of a wave packet}

Numerical solutions (not here described in detail) suggest that there are cases in which the wave packets organize the vortices into patterns. As a first step in understanding this phenomenon we study the stability of a vortex street in the presence of a single wave packet \textcolor{black}{(per period)} in the semi-periodic domain.
This system conserves energy and momentum, but angular momentum is not conserved because the periodicity breaks rotational symmetry. 

\textcolor{black}{Within the (periodic) domain}, we have four vortices, arranged symmetrically about  $y=0$, with $y=\pm b$, and spaced $\pi$ apart in the $x$-direction.  Refer to figure 8. This is the minimal system to illustrate the instability of a periodic vortex street \citep{domm1956}. The stability of vortex streets was first considered by \citet{vonkarman1911}, and discussed in detail by \citet[\S 156]{Lamb1932}. \citet{domm1956} examined its nonlinear stability, reducing the system of four vortices to two dependent variables. The \textit{symmetric} vortex street proved to be linearly unstable in all of parameter space, whereas the asymmetric vortex street is linearly stable at a single value of the ratio of vertical to horizontal spacing of the vortices.  However, even the linearly stable asymmetric vortex street is nonlinearly unstable \citep{domm1956}.

\begin{figure} 
  \centering
  \includegraphics[width=3.0in]{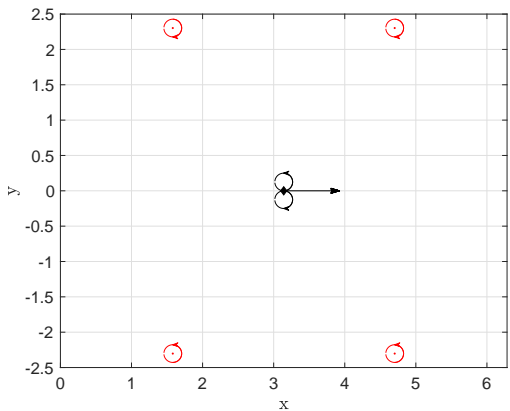}
  \caption{The vortex street configuration considered in this section. The vortices in the top row have strength $\Gamma$ while those in the bottom row have strength $-\Gamma$. The domain is periodic in the $x$-direction, and there is a wave packet travelling along the $x$-axis. We study the stability of this vortex street in the presence of the wave packet.   }
  \label{fig:stability}
\end{figure}

The equations of motion are (\ref{047}) - (\ref{049}) with the stream functions given by (\ref{141}) and (\ref{142}).
We take $\Gamma_i = \epsilon^2 \gamma$ so that $\Gamma_1 = \Gamma_2 = \epsilon^2\gamma$ and $\Gamma_3 = \Gamma_4 = - \epsilon^2\gamma$.
We seek equations of motion valid to $O(\epsilon^2)$. We begin by expanding the variables describing the wave packet as
\beq
x_p = x_{p0}+c_{g0}t+\epsilon^2 x_{p2},\quad y_p = 0,
\label{173}
\eeq
\beq
k_p = k_{p0},\quad l_p=0.
\label{174}
\eeq
If there are no second order corrections to the wavenumbers initially, and if the momentum is conserved, then the second order wavenumbers must remain zero for all time. 

The governing second-order equation for the horizontal motion of the wave packet is
\beq
\dot{x}_{p2} = -\frac{1}{4\pi} \sum_i \gamma_i \frac{\sinh y_{i0}}{\cos \chi_{i0}-\cosh y_{i0}}.
\label{175}
\eeq
Expanding hyperbolic functions, we find that
\beq
\dot{x}_{p2} = \frac{\gamma}{ \pi }+O(\epsilon),
\label{176}
\eeq
where $\gamma \equiv \gamma_1$. Eqn (\ref{176}) implies a correction to the wave packet speed, due to the velocity induced by the vortex street, which has the magnitude of the mean velocity in the plane of symmetry of the vortex street \citep[\S 156]{Lamb1932}. Expanding
\beq
\dot{y}_{p2} = \frac{1}{4\pi}\sum_i \frac{\sin \chi_{i0} }{\cos \chi_{i0}-\cosh y_{i0}} = 0+O(\epsilon),
\label{177}
\eeq
we find that there is no correction to the vertical speed of the wave packet. 

We now solve for the motion of the point vortices. It will be a combination of the motion induced by the wave packet (as calculated in \S 6) and the uniform self advection of the vortex street. 
\citet[\S 156]{Lamb1932} finds that the self advection of the vortex street leads to a uniform translation of the vortices with speed
\beq
\epsilon^2 \frac{\gamma}{2\pi} \coth b =  \epsilon^2 \frac{\gamma}{2\pi} + O(\epsilon^3). 
\label{178}
\eeq
Then, from (\ref{172}), we obtain the second order behavior of the vortices as 
\beq
z_{2i} = \left(\frac{\gamma}{2\pi} +  \frac{\mu^2}{c_{g0}}\right)t-\frac{\mu }{ c_{g0}}\left(e^{-2\ii\chi_{i0}  }-1\right) + \frac{ \mu^2}{c_{g0}^2}( e^{-\ii \chi_{i0}}-1),
\label{179}
\eeq
where, from our initial conditions $\chi_{10} =\chi_{30}= -\pi/2 -c_{g0}t$ and $\chi_{20}=\chi_{40}=\pi/2 - c_{g0}t$.
Note the presence of two mean flows. One is induced by the wave packet; the other represents the self advection of the vortex street. These two competing mean flows are now shown to have implications for the stability of the vortex street. 

\subsection{Linear stability analysis}
Now we perturb our system to examine its linear stability. We expand the vortex locations as
\beq
z_i = z_{i0}+\epsilon z_{i1}+\epsilon^2 z_{i2} + \delta z_{\delta i},
\label{194}
\eeq
\beq
\quad x_p = x_{p0}+\epsilon^2 x_{p2}+\delta x_{p\delta}, \quad y_p=\delta  y_{p\delta},\quad  k_p = k_{p0}+\epsilon^2\delta k_{p\delta}, \quad l_p =\epsilon^2\delta l_{p\delta},
\label{195}
\eeq
where $\delta$ is the amplitude of the perturbations. The goal is to expand the equations of motion to $O(\epsilon^2 \delta)$.
Starting with the motion of the wave packet, and using the results found in the previous subsection, we find that at $O(\delta)$
\beq
\dot{x}_{p\delta} = c_{gx\delta},
\label{196}
\eeq
\beq
\dot{y}_{p\delta} = c_{gy\delta},
\label{197}
\eeq
where $(c_{gx\delta},c_{gy\delta})$ are the $O(\delta)$ expansions of the group velocity.
The wavenumbers evolve according to 
\beq
\dot{k}_{p\delta}=\dot{l}_{p\delta}=0.
\label{198}
\eeq
If our initial conditions are such that these perturbation wavenumbers vanish, they will vanish for all time. This implies that $(x_{p\delta},y_{p\delta})$ are constants which we take to be zero. 

The nontrivial part of the analysis comes from the evolution of the vortices.  We need $\bold{U}_m$ and $\bold{U}_d$ to $O(\epsilon^2 \delta)$.
At this order, the vortex induced flow is the same as it would be in the absence of the wave packet, hence, from \citet{Lamb1932}, we have the $O(\epsilon^2 \delta)$ result
\begin{align}
&\bold{U}_{m}(z_1) = -\ii\frac{\gamma}{8\pi}(- z_{\delta 1}+z_{\delta 2}),\label{199} \\
&\bold{U}_{m}(z_2) = \ii\frac{\gamma}{8\pi}(- z_{\delta 1}+z_{\delta 2}).
\label{200}
\end{align}
At this order $\bold{U}_m(z_1)$ and $\bold{U}_m(z_2)$ only depend on vortex 1 and 2, hence we omit $\bold{U}_m(z_3)$ and $\bold{U}_m(z_4)$ for clarity of presentation. That is, we only need to consider the evolution of vortices 1 and 2, or equivalently vortices 3 and 4. 

The $O(\delta)$ contributions to $\bold{U}_d(z_{1,2})$ are found by substituting the expansion given by (\ref{194}) into (\ref{160}) so that
\beq
\bold{U}_d(z_i) = \mu \epsilon \delta e^{-2\ii \chi_{i0}}z_{\delta i}^* (\ii e^{\ii \chi_{i0}}+\epsilon \left(4\ii +e^{\ii \chi_{i0}}z_{i1}^*)\right).
\label{201}
\eeq
To solve this system of equations, we must also expand the perturbations in $\epsilon$, so that
\beq
z_{\delta i} = z_{\delta i0} +\epsilon z_{\delta i1}+\epsilon^2 z_{\delta i2}.
\label{202}
\eeq
We assume the time dependence of the order zero terms goes like $e^{\Lambda t}$, where $\Lambda=\epsilon^2\lambda $ (which may be inferred from the classical stability analysis, which implies that the growth rates are proportional to the strength of the vortices, see \citealt{Lamb1932}). Additionally, we assume that the first order terms have fast oscillations, so that their time derivatives have the same order as the original term. We need not solve explicitly for the second order terms to conduct the stability analysis. 

The restrictions on the first order terms are found to imply
\beq
z_{\delta 1 1} = \ii\frac{\mu}{2c_{g0}}z_{\delta 10}^*(e^{\ii \theta}-1) ,
\label{203}
\eeq
and
\beq
z_{\delta 2 1} =  -\ii\frac{\mu}{2c_{g0}}z_{\delta 20}^*(e^{\ii \theta}-1).
\label{204}
\eeq
We now have sufficient information to solve for the stability of the vortex street configuration, which will be dictated by the evolution of $\{z_{\delta 10},z_{\delta 20}\}$. Substituting the lower order expansions into $\bold{U}_d$ and $\bold{U}_m$, we find that the dynamics of $\{z_{\delta 10},z_{\delta 20}\}$ are given by the phase averaged equations of motion. 

The evolution equations are given by
\begin{align}
&\dot{z}_{\delta 1 0} =  \epsilon^2\left(-\frac{\ii \nu}{2}(z_{\delta 10}^*-z_{\delta 20}^*)+\frac{\ii \sigma }{2}(z_{\delta 10}-z_{\delta 10}^*)  \right),\label{205}\\
&\dot{z}_{\delta 2 0} =\epsilon^2\left(\frac{\ii \nu}{2}(z_{\delta 10}^*-z_{\delta 20}^*)+\frac{\ii \sigma }{2}(z_{\delta 20}-z_{\delta 20}^*)  \right),
\label{206}
\end{align}
where
\beq
\sigma = \frac{\mu^2}{2c_{g0}}; \quad \nu = \frac{\gamma}{4\pi}.
\eeq
Define $z_{\delta 10} = \alpha e^{ \epsilon^2 \lambda t},z_{\delta 20} = \beta e^{ \epsilon^2 \lambda t}$. Then, (\ref{205}-\ref{206}), together with their complex conjugates, imply the following eigenvalue problem 
\beq
\begin{pmatrix}
\sigma-2\ii\lambda  & 0 & -\nu-\sigma & \nu \\
0 & \sigma-2\ii\lambda & \nu & -\nu-\sigma \\
\nu+\sigma &-\nu &-\sigma-2\ii\lambda & 0 \\
-\nu &\nu+\sigma & 0 & -\sigma-2\ii\lambda  \\ 
\end{pmatrix}
\begin{pmatrix}
\alpha\\
\beta\\
\alpha^*\\
\beta^*\\ 
\end{pmatrix}
=\bold{0}
\eeq 
The eigenvalues $\lambda$ are found to be
\beq
\lambda^2 =\nu(\nu+\sigma).
\eeq
In the limit that $\sigma=0$, we recover the result of \citet[\S 156]{Lamb1932} that the symmetric vortex street is linearly unstable.  It follows from (7.23) that the system is stable when 
\beq
\nu(\nu+\sigma)<0.
\eeq
We note that $\nu$ can be positive or negative depending on the sign of $\gamma$, so that the inequality may hold only if the sign of $\nu$ and $\sigma$ are different. Physically, the stability depends on the sign of the induced motion of the vortices. If the drift induced by the wave packets is larger than the self advection of the vortices, \textcolor{black}{and $\nu<0$}, the system remains linearly stable.

\subsection{Long-time behavior}
In the previous subsection, we analyzed the linear stability of the vortex street to perturbations. However, as was shown by \citet{domm1956}, even a linearly stable vortex street might be nonlinearly unstable.
This may be examined analytically, but the algebra becomes considerably involved, so we instead perform a numerical investigation. We integrate our equations of motion using a fourth order Runge-Kutta scheme, ensuring that the Hamiltonian and momentum are conserved. We also increase the number of adjacent \textcolor{black}{periodic extensions} until convergence is found \textcolor{black}{(here we take our total domain to be of length 5001$\times 2\pi$)}. We take $\epsilon = 1/4, k_0 = 1, \mathcal{A}_p=1, \Gamma = \epsilon^2$, and we integrate the system for $3\times 10^4$ seconds. This is shown in the left panel of figure 9, where it is seen that the vortex street remains stable for long times in this configuration. In the absence of the wave packet $(\mathcal{A}_p=0)$, shown on the right panel of figure 9, the system is unstable, and the vortex street eventually dissolves, analogous to the behaviour found for two vortex pairs \citep{Love1893,Tophoj2013}.

\begin{figure} 
  \centering
  \includegraphics[width=5in]{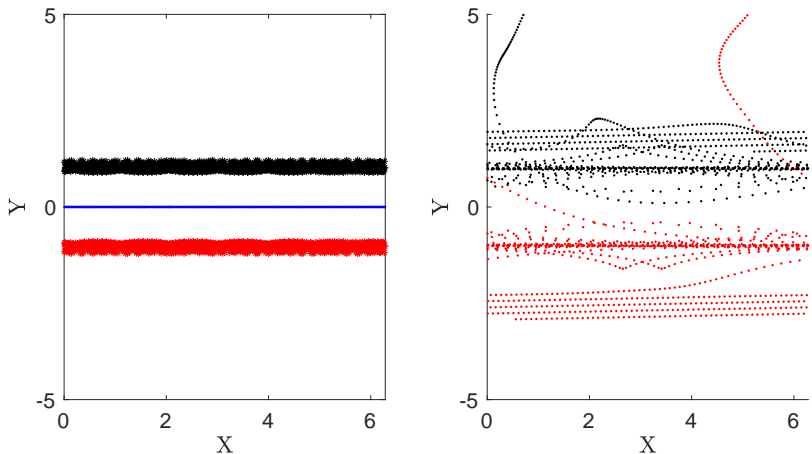}
  \caption{Left: Stability of periodic (in the x-direction) vortex street, with initial conditions depicted in figure 8. The wave packet motion is shown in blue, while the vortices are shown in black ($\Gamma_i<0$) and red $(\Gamma_i>0)$. The left panel shows that the vortex street is stable over this long time integration. The right panel shows the same initial configuration of the vortices with no wave packet, and we see the system is unstable, with vortices propagating far away from their initial locations.  }
  \label{fig:stability}
\end{figure}

\section{Conclusion}

It has long been recognized in the oceanographic community that surface waves may freely exchange momentum and energy with underlying currents. The dynamics are governed by wave action conservation \citep{LHS1962, whitham1965}, while the evolution of the phase obeys the equations of geometrical optics. These equations are valid for small amplitude inviscid waves that are slowly varying. Despite the maturity of this theory \citep{Phillips1966}, there are still many fundamental questions regarding wave - current interaction, including a need to better understand their two way coupling \citep{McWilliams2004}. This is thrown in to particularly stark relief by numerical models of climate, which are beginning to resolve the submesoscale (on the order of 1-10km), where these interactions may be especially pronounced \citep{McWilliams2016, Romero2017}. At even smaller scales, wave breaking in deep-water occurs for waves with finite crest length, which implies that at the free surface the breaking induced flow is characterized by a dipole structure \citep{peregrine1999, Pizzo2013}. The interaction of this flow with the wave field is thought to be significant for establishing Langmuir circulations \citep{Leibovich1983}, a crucial process for mixing the upper ocean. However, these classical theories do not take into account finite bandwidth effects, nor do they account for the two way coupling between the wave and current fields.

Recently, there have been efforts to better describe two-way coupling effects \citep[e.g.][]{Phillips2002, McWilliams2004, Suzuki2019}. Although these theories provide crucial insight, they are complicated and often obscure simple underlying physical constraints. Here, we have provided a simplified framework to examine wave-current interaction by assuming that the wave packets are compact, and that the currents are a collection of point vortices. Since this simplified system is derived from a variational principle, conservation laws arise naturally from the symmetries of the Lagrangian.

The central assumption made in this study is that the Doppler-shifted dispersion relationship serves as a faithful starting point to model wave - current interaction. That is, no additional terms are needed in the dispersion relationship to account for the vortical nature of the currents (see, for example, \citet{Stewart1974} to see how vertical shear may modulate the dispersion relationship). Additionally, we assume that the currents (in the form of point vortices) and the wave packets are compact and widely spaced. A further simplifying but unnecessary assumption is that the wave packets do not interact with each other.

We have examined several solutions for the case of one wave packet and one vortex. These include stable bound orbits and unstable configurations. The wave packet and vortex may capture one another. We have also examined situations in which the wave packet and vortex collapse, occupying the same location at the same time. When collapse occurs, our theory breaks down. We also considered blow-up solutions, in which the modulus of the wavenumber grows without bound. In reality this growth would be arrested by wave breaking.  

After examining a solution with two point vortices and one wave packet, we considered the motion induced by a wave packet on a weak point vortex in a horizontally periodic domain. It was shown that a net drift is induced by the wave packet, which may have possible implications for the advection of jetsam, flotsam and pollution at the ocean surface. This drift also has implications for the stability of a vortex street in the presence of a wave packet. Our analysis shows that the wave packet may stabilize the vortex street. Numerical calculations confirm that this system is stable for long times, suggesting that this phenomenon might be observable in nature. 

This work motivates several future studies. In particular, the addition of the generation of waves by wind, wave packet - wave packet interaction, and wave breaking, which creates a pair of oppositely signed vortices, would make this a more realistic description of the upper ocean.   \\

Declaration of Interests: The authors report no conflicts of interest.\\

\appendix
\textbf{Appendix A.  Justification of Whitam's Lagrangian}
\bigskip

This Appendix offers a derivation of \eqref{01} following Whitham's averaged-Lagrangian method.  Our starting point is the linear approximation to the Lagrangian of \cite{Miles1977}---see also \citet{luke1967} and \citet{Zakharov1968}---namely
\beq
L[\phi,\eta]=\int dt \int dx \left( \phi \eta_t - H[\phi,\eta] \right),
\label{A1}
\eeq
where
\beq
H[\phi,\eta]= \frac{1}{2}g\eta^2 + \int_{-\infty}^0 dz \; 
\frac{1}{2}\left( \phi_x^2+\phi_z^2 \right).
\label{A2}
\eeq
Here $\phi$ is the velocity potential and $\eta$ is the surface elevation.
The integral in \eqref{A1} is over the sea surface $z=0$, and, for simplicity of notation, we ignore the $y$-direction.
Variations $\delta \phi$, $\delta \eta$ yield the familiar linear equations and boundary conditions.  A solution is
\beq
\eta(x,t)=A\cos (kx-\omega t), \;\;\;\;\; 
\phi(x,z,t)=\frac{A\omega}{k}e^{kz}\sin (kx-\omega t),
\label{A3}
\eeq
where $A$ and $k$ are constants, and $\omega^2=gk$. Following Whitham, we substitute
\beq
\eta(x,t)=A(x,t)\cos (\theta(x,t)), \;\;\;\;\; 
\phi(x,z,t)=\frac{A(x,t)\theta_t}{\theta_x}e^{\theta_x z}\sin (\theta(x,t))
\label{A4}
\eeq
back into \eqref{A1} and \eqref{A2}, obtaining
\begin{align}
L[A,\theta]&=\iint dtdx \; \frac{\omega^2 A^2 }{k}\sin^2\theta
-\int dt \; H[A,\theta]
\notag \\
&=\frac{1}{2}\iint dtdx \; \frac{\omega^2 A^2 }{k}
-\int dt \; H[A,\theta]
\label{A5}
\end{align}
and
\begin{align}
H[A,\theta]&=\int dx \; A^2 \left( 
\frac{1}{2}g\cos^2\theta
+\frac{1}{4}\frac{\omega^2}{k}\cos^2\theta
+\frac{1}{4}\frac{\omega^2}{k}\sin^2\theta \right)
\notag \\
&= \frac{1}{4}\int dx \; A^2 \left( g + \frac{\omega^2}{k} \right),
\label{A6}
\end{align}
where now $\omega=-\theta_t$ and $k=\theta_x$.  In the final step of \eqref{A5} and \eqref{A6}, we average over the fast dependence of $\theta(x,t)$.  Combining results, we have
\beq
L[A,\theta]= \frac{1}{4}\iint dtdx \left( \frac{\omega^2}{k}-g \right)A^2
=\frac{1}{2}\iint dtdx \left( \omega-\frac{gk}{\omega} \right)  \mathcal{A},
\label{A7}
\eeq
where, by \eqref{A6},
\beq
E=\frac{1}{2} gA^2=\frac{1}{2} \frac{\omega^2}{k}A^2
\label{A8}
\eeq
is the wave energy per unit area, and $\mathcal{A}=E/\omega$ is the action.  Independent variations of $A$ and $\theta$ are equivalent to independent variations of $\mathcal{A}$ and $\theta$, and either choice of variation yields the dispersion relation and the action conservation equation for surface waves.  The Lagrangian
\beq
L[\mathcal{A},\theta]=\iint dtdx \left( \omega-\sqrt{gk} \right) \mathcal{A}
\label{A9}
\eeq
yields the same two equations and is equivalent to \eqref{01}.

\bigskip
\noindent
\textbf{Appendix B.  Stream function generated by a single wave packet.}
\bigskip

Let the wave packet be located at $\mathbf{x}=0$.    For $r\equiv \vert \mathbf{x} \vert \gg \vert \mathbf{x}' \vert $,
\beq
\ln  \vert \mathbf{x} - \mathbf{x}' \vert \approx \ln r - \frac {\mathbf{x} \cdot \mathbf{x}'}{r^2}
\label{B1}
\eeq
and \eqref{033} becomes
\beq
\psi_w(\mathbf{x})\approx \frac{\ln r}{2 \pi} \iint d\mathbf{x}' \; \rho(\mathbf{x})
- \frac{1}{2 \pi} \frac {\mathbf{x}}{r^2} \cdot  
\iint d\mathbf{x}' \;  \mathbf{x}' \rho(\mathbf{x}).
\label{B2}
\eeq
The first term in \eqref{B2} vanishes, because $\mathcal{A}=0$ at the boundary of the wave packet. In the second term,
\beq
\iint d\mathbf{x} \;  \mathbf{x} \rho(\mathbf{x}) 
=\iint d\mathbf{x} \;  \mathbf{x} \left( \mathcal{A}_x l_p -\mathcal{A}_y k_p \right)
= \mathcal{A}_p (-l_p,k_p)
\label{B3}
\eeq
after integrations by parts, where
\beq
\mathcal{A}_p = \iint d\mathbf{x} \mathcal{A}.
\label{B4}
\eeq

\bigskip
\noindent
\textbf{Appendix C.  Elimination of $\psi$ from the Lagrangian}
\bigskip

In this Appendix we justify the step of using the equations obtained by varying a particular field $\psi(\mathbf{x},t)$ to eliminate that same field from a Lagrangian that depends on several fields.  We show that variations of the modified Lagrangian yield the correct equations for the remaining fields.

First we consider the related problem of finding the stationary points---maxima, minima or inflection points---of an ordinary function of two variables.  Let the function be $f(\phi,\psi)$.  The stationary points are found by solving the set
\beq
\frac{\partial}{\partial \phi} f(\phi,\psi)=f_1(\phi,\psi)=0
\label{C1}
\eeq
and
\beq
\frac{\partial}{\partial \psi} f(\phi,\psi)=f_2(\phi,\psi)=0
\label{C2}
\eeq
where $f_1$ denotes the derivative of $f$ with respect to its first argument, and $f_2$ denotes the derivative of $f$ with respect to its second argument.  Suppose that (\ref{C2}) can be solved explicitly for $\psi$ in the form
\beq
\psi=g(\phi)
\label{C3}
\eeq
Then, substituting (\ref{C3}) into (\ref{C1}) we obtain a single equation for $\phi$, namely
\beq
f_1(\phi,g(\phi))=0
\label{C4}
\eeq
Our contention is that (\ref{C4}) is equivalent to
\beq
\frac{\partial}{\partial \phi} f(\phi,g(\phi))=0
\label{C5}
\eeq
Clearly (\ref{C5}) is equivalent to
\beq
f_1(\phi,g(\phi))+f_2(\phi,g(\phi))g'(\phi)=0
\eeq
and the fact that (\ref{C3}) solves (\ref{C2}) means that $f_2(\phi,g(\phi))=0$.  Thus (\ref{C5}) is indeed equivalent to (\ref{C1}).

To see that this proves our contention about the Lagrangian, replace the integral over space and time by a sum over gridded values, and replace the derivatives of the field variables by finite differences. Then the Lagrangian becomes an ordinary function of many variables, namely, the gridded values of the fields.  We again regard this function as $f(\phi,\psi)$ where now $\psi$ stands for a vector whose components are all the gridded values of $\psi$, and $\phi$ stands for a vector whose components are all the gridded values of $\phi$.  If there are $N$ spacetime gridpoints, then (\ref{C1}) and (\ref{C2}) each represent $N$ equations, but the essence of the proof is the same as that given above.  It is easy to invent examples that show that the use of (\ref{C3}) to eliminate some \emph{but not all} of the $\psi$-terms in $f(\phi,\psi)$ leads to erroneous results.  The results given here border on the trivial, but the strategy of completely eliminating a field using the equations that result from the variations of that same field is important, because it seems to be one of the few legitimate methods of using the results of a variational principle to simplify the variational principle itself.

\bibliography{ref}
\bibliographystyle{jfm}

\end{document}